# Techno–Economic Modeling and Safe Operational Optimization of Multi-Network Constrained Integrated Community Energy Systems

Ze Hu[a], Ka Wing Chan[a,b], Ziqing Zhu[c*], Xiang Wei[a], Weiye Zheng [d], Siqi Bu[a,b,e]

[a] Department of Electrical and Electronic Engineering, The Hong Kong Polytechnic University, Hung Hom, Hong Kong SAR
[b] Research Centre for Grid Modernisation, The Hong Kong Polytechnic University, Hung Hom, Hong Kong SAR
[c] Department of Computer Science, The Hong Kong Baptist University, Kowloon Tong, Hong Kong SAR
[d] School of Electrical Power Engineering, South China University of Technology, Guangzhou, China
[e] International Centre of Urban Energy Nexus, The Hong Kong Polytechnic University, Hung Hom, Hong Kong SAR

**Abstract:** The integrated community energy system (ICES) has emerged as a promising solution for enhancing the efficiency of the distribution system by effectively coordinating multiple energy sources. However, the concept and modeling of ICES still remain unclear, and operational optimization of ICES is hindered by the physical constraints of heterogeneous integrated energy networks. This paper, therefore, provides an overview of the state-of-the-art concepts for techno–economic modeling of ICES by establishing a Multi-Network Constrained ICES (MNC-ICES) model. The proposed model underscores the diverse energy devices at community and consumer levels and multiple networks for power, gas, and heat in a privacy-protection manner, providing a basis for practical network-constrained community operation tools. The corresponding operational optimization in the proposed model is formulated into a constrained Markov decision process (C-MDP) and solved by a Safe Reinforcement Learning (RL) approach. A novel Safe RL algorithm, Primal-Dual Twin Delayed Deep Deterministic Policy Gradient (PD-TD3), is developed to solve the C-MDP. By optimizing operations and maintaining network safety simultaneously, the proposed PD-TD3 method provides a solid backup for the ICESO and has great potential in real-world implementation. The non-convex modeling of MNC-ICES and the optimization performance of PD-TD3 is demonstrated in various scenarios. Compared with benchmark approaches, the proposed algorithm merits training speed, higher operational profits, and lower violations of multi-network constraints. Potential beneficiaries of this work include ICES operators and residents who could be benefited from improved ICES operation efficiency, as well as reinforcement learning researchers and practitioners who could be inspired for safe RL applications in real-world industry.

## 1. Introduction

### 1.1 Motivation

With the increasing fossil fuel depletion and climate change issues, the energy system is significantly transitioning worldwide [1]. Increasing energy efficiency and renewable utilization to realize decarbonization is the main aim of this transition and has enormous meaning to the whole human society. On the other hand, the advent of the second technological revolution has seen a predominant reliance on electrical energy as the main source to partially replace and supplement traditional fossil fuels for the deployment and operation of energy systems. The multiple energy resources, including electric energy and fuels, are transmitted with different means and speeds to the demand side, undergoing specific conversions to thermal, kinetic, and light energy to meet specific applications. From this point of view, multi-energy coordination can change the whole picture of the energy system and thus promoting the energy transition and decarbonization by affecting the operational logic due to the inherent interconnection of multiple energies [2].

Notably, community buildings, which consumed 32% of primary energy in the U.S. in 2019—13% in the commercial buildings sector and 19% in the residential sector, highlight the substantial potential for energy integration on the demand side [3]. Furthermore, the proliferation of distributed energy devices and the evolution of energy distribution networks lay a solid foundation for such integration, which can assist in improving the system profits by operational optimization. Therefore, energy integration and operational optimization on the demand side are of great significance in promoting energy transition and are getting much attention. It is necessary and pressing to increase energy efficiency and utilization with the means of energy integration in the whole system.

Integrated Community Energy Systems (ICES) have emerged as a promising approach for efficient multi-energy coordination and utilization, particularly in managing demand flexibility and increasing renewable energy penetration [4]. The operational optimization of ICES is, therefore, essential for integrating diverse energy transactions and enhancing overall energy efficiency. However, the concept of ICES, which represents an integrated energy system at the community level, is still under discussion and lacks a clear definition. Some researchers have described ICES as a modern development that reorganizes local energy systems to integrate distributed energy resources and engage local communities [5]. Others have focused on its role in managing local energy generation, delivery, and exchange to meet local demand, with or without grid connection [6]. However, these descriptions do not fully capture the operational logic and model structure of ICES. Inspired by the concept of energy communities [7, 8], we define *ICES* as follows: ICES is a socio-economic unit rooted in a physical community, characterized by cooperative multi-energy production and consumption through either shared or unshared integrated energy devices, and functioning as a non-commercial market actor that amalgamates economic, environmental, and social community objectives. While sharing the goal of maximizing social welfare through energy device scheduling and demand response stimulation, ICES extends beyond electric energy to include the integration of power, gas, and heat, emphasizing coordination among both energy devices and demands. Thus, ICES represents an effective strategy for maximizing social welfare and facilitating decarbonization.

### 1.2 Related work

Although the concept of ICES has not been explicitly defined in prior research, a substantial body of work has explored ICES in multiple aspects, including system modeling, operational problems, and optimization approaches [9-18]. Therefore, related works are mainly reviewed in terms of these aspects.

**System Modeling in ICES**: Based on a review of previous studies, the modeling of ICES can be divided into three main components: community-level devices, consumer-level devices, and network constraints. Devices in ICES can be divided into community-level and consumer-level. Community-level devices typically include dispatchable generation (DG) units, energy storage systems (ESS), and renewable energy sources (RES). DG units comprise combined heat and power (CHP) systems [12,14-17], power-only units [12,15,16], and heat-only units [12,15-17]. CHP systems, which serve as critical energy converters across power, gas, and heat, are modeled simplistically with fixed energy conversion rates in most works of ICES [12,14-17]. However, the realistic and physical characteristics of CHP are always overlooked, which describes the multi-energy conversion as a feasible operation region (FOR) but presents computational challenges with non-convexity [19]. Power-only and heat-only units are rarely used, which are typically modeled using a linear [12,14-17] or quadratic generation cost function [13,16]. For ESS, electric battery systems (EBS) [8-15], thermal energy storage (TES) [8-13], and gas storage systems (GSS) are considered. Compared to prevalent EBS and TES, which have variable costs, GSS is less common due to static gas prices. Typical simplified ESS models are usually employed with static value for charging and discharging efficiency. This is because ESS does not directly participate in multi-energy conversion and is the core part of the ICES, although ESS is deemed necessary. RES, such as photovoltaic (PV) systems and wind turbines (WT), introduce renewable power output with uncertainty, constituted of an energy conversion model and forecast errors. The energy conversion model provides the output given the solar irradiation or wind speed and other external conditions (e.g., temperature [11]), while the forecast errors are sampled from specific probabilistic distribution functions (PDFs). For example, Weibull and Beta PDFs can be used to represent the forecast error distribution for the WT and PV, respectively [9]. At the consumer level, the modeling focuses on the energy demand for electricity and heat, typically based on a quadratic energy utility function to represent demand response characteristics. Households may possess energy conversion or flexible devices like micro-CHP, ESS, and boilers, sometimes overlapping with community-level devices. The boilers employed on the demand side enable energy conversion to realize a more flexible integrated demand response (IDR) [17]. Moreover, some studies have extended ICES modeling to include more detailed flexible devices, for example, electric vehicles (EVs) [9], to explore the unique characteristics of ICES in various scenarios. Additionally, as research accounting for network constraints in ICES is very limited, previous works in network modeling are reviewed in the following operational research part rather than separately.

**Operational Problems on ICES**: The existing literature on the operational optimization of ICES primarily addresses the coordination of two energy systems, a focus partly due to the complexity and computational intensity involved. For power and heat systems within ICES, prior research has primarily concentrated on leveraging thermal demand characteristics, given their direct impact on human comfort. IDR strategies for power and heat have been employed to manage uncertainty and enhance profitability without compromising comfort levels. For instance, in [9], the coordination between flexible IDR for power and heat and electric vehicle charging stations (EVCS) is explored in ICES under the uncertainty of renewable generation. A bi-level model predictive control (MPC) based approach is utilized in [16] to optimally integrate thermal demand and flow dynamics into the ICES scheduling problem. [17] optimizes the distributed scheduling problem of multiple energy hubs in ICES. Furthermore, [18] considers the impact of the thermal inertia of detailed space heating loads to model the thermal demand response character in the IDR problem of multiple energy users (MEUs) in ICES. On the other hand, literature on power and gas systems in ICES is limited due to the non-convex nature of gas flow, focusing on the coordination of energy flow in distribution networks. While research on two-network coupling systems is extensive, the coordination and interaction among multiple networks are still underexplored. Notably, comprehensive modeling and mathematical optimization of multi-networks are proposed in a multi-energy district [20], which shares a similar scale with ICES. However, multi-energy districts primarily concentrate on network operations without addressing energy device scheduling and are considered centrally controlled entities, in contrast to the community-oriented nature of ICES. As a result, the multi-network constrained scheduling of ICES operators and the interaction (e.g., IDR) between ICES operators and MEUs remain critical yet underexplored aspects—the modeling and operational optimization of multi-network constrained ICES warrants further investigation.

**Optimization Approaches of ICES**: The constrained optimization problem in ICES is challenging to solve in terms of non-convexity, privacy protection, and computational burden, which are caused by non-convex constraints of devices and network, the distributed operation manner of MEUs, and the increasing scale of the modern community, respectively. It can be solved by multiple approaches, including heuristic algorithms [14] and mathematical programming [15-18]. Heuristic algorithms are a class of optimization algorithms that are designed to explore solution spaces to find near-optimal solutions efficiently. Therefore, this approach is particularly useful for problems with non-convexity. A metaheuristic algorithm, chaotic differential evolution (CDE), is adopted to schedule and price for multiple ICESs in [14]. However, it fails to guarantee optimality theoretically and is easy to fall into suboptimal, especially in large-scale and complex problems like ICES operation. In contrast, mathematical programming guarantees the solution optimality with rigorous proof but falls short in dealing with non-convexity, which requires complicated convexification. Works in [15] formulate a convex problem by employing the Big-M method and inequality second-order cone (SOC) constraint, after which active and reactive dispatching for ICES is solved to the global optimal. Similarly, network constraint nonlinearity was tackled using the big-M method and piece-wise linearization [18]. The two-stage optimization of the power and heat system in ICES is then solved by a robust method subject to energy price uncertainty. However, these approaches above require global information for solutions, violating the privacy protection of MEUs in an ICES. To overcome this drawback, the alternating direction method of multipliers (ADMM), was adopted to schedule the sub-systems of ICES in a decentralized manner [17]. Even though these approaches can partially deal with non-convexity and realize privacy protection, they still face increasing computational burdens with the growing scalability of the consumers and devices, which is known as the "curse of dimensions [21]."

***Software Programs for ICES***: Meanwhile, multiple software programs are designed for operating and planning in various levels of energy systems. At the transmission level (nation or province), the Integrated Energy Management System (IEMS) [22] developed by Tsinghua University focuses on solving the multi-energy flow and optimization in bulk energy systems. EnergyPLAN [23] of Aalborg University is to assist in the design of national energy planning strategies with technical and economic analyses of the consequences of different choices and investments. At the distribution level, the Hybrid Optimization Model for Electric Renewables (HOMER) software [24] is developed by the National Renewable Energy Laboratory (NREL) for multi-energy microgrid planning. It provides financial analysis on user predefined combinations of technologies and sensitivity analyses given external uncertainty rather than optimizing decisions. In contrast, the Distributed Energy Resources Customer Adoption Model (DER-CAM) [25] designed by Lawrence Berkeley National Laboratory, Energy Efficiency and Risk Management in Public Buildings (EnRiMa) [26] of the European Union, and Toolkit for Optimization of Industrial Energy Systems (TOP-ENERGY) [27] from RWTH Aachen University, Planning and Designing of Microgrid (PDMG) [28] designed by the Tianjin University, the CEPAS developed by Xiamen University [29], are optimization-based programs. This sort of software program provides optimal day-ahead operational orders and long-term strategic planning solutions by minimizing the energy system operational costs in a microgrid, public building, or factory. Specifically, DER-CAM is supposed to find a static optimal solution for equipment combination and its operation over a typical year (average over many historical years) in an electricity-heating system. EnRiMa, however, enables multi-objective optimization by lowering costs with given comfort and financial risks. It also considers long-term perspective with multi-stage stochastic scenario trees, improving the robustness of its solution. TOP-ENERGY focuses on optimizing the operation of industry energy supply systems. A hybrid method combining mathematical programming and a heuristic algorithm is adopted to improve the solution's accuracy and performance. PDMG incorporates a more detailed energy device model (e.g., battery and supercapacitor) in the system planning than other software. CEPAS has the advantage of adopting multi-objective optimization and multi-agent distribution models for regional IES planning and operation. It also enables agent-based energy demand forecasting and can deal with non-linearity. Most of the software programs above are rule-based or largely rely on mathematical programming, which somehow falls short in dealing with non-convexity, scalability, and privacy protection.

***Reinforcement Learning Approach***: In recent years, Reinforcement Learning (RL) algorithms have gained great attention for addressing optimization problems [30]. By interacting with the external environment, RL algorithms enable intelligent agents to iteratively learn optimal strategies with only partial environmental information. Compared to traditional mathematical programming, RL offers advantages in scalability with high computational efficiency and generalization to various scenarios [31]. Moreover, Deep RL (DRL) algorithms, leveraging Deep Neural Networks (DNN) to estimate value functions, can handle complicated optimization problems by estimating non-convex Q-functions or policies. DRL has been successfully applied in diverse domains [19, 32-39]. For example, model-free DRL algorithm, Deep Deterministic Policy Gradient (DDPG), optimizes the energy management of an integrated energy hub in [19]. Similarly, Soft Actor-Critic (SAC) algorithms optimize the scheduling of islanded energy systems, accounting for multi-uncertainties and hydrothermal simultaneous transmission [32]. RL algorithms can also be extended to multi-agent environment in peer-to-peer multi-energy trading [40, 41], showing their strong adaptability to different settings. However, conventional RL algorithms are designed for unconstrained optimization problems. Even though they are applicable to some optimization problems with soft constraints, their efficacy may diminish when applied to most constrained optimization problems. The lack of consideration for network constraint violations restricts the application of DRL algorithms in industrial practice, as it can lead to economic losses and even system blackouts [33].

***Safe Reinforcement Learning***: To address this challenge, Safe RL algorithms have been developed to solve constrained optimization problems, which are designed to maximize reward while complying with hard constraints. Specifically, Safe RL approaches can be classified into three categories: 1) Penalizing constraint violations in the reward function by adding a penalty term [34]. However, this requires choosing a suitable penalty value, which is a difficult and sensitive task that depends on the reward scale, the number and scale of constraints, and the degree of safety [35]. 2) Projecting unsafe actions to safe ones by solving a projection problem, for example, approximated Lyapunov constraints [36]. This method relies on a projection model, which is based on predefined DNNs or matrices with potentially large approximation errors [37]. Therefore, the resulting actions could be overly conservative. 3) Penalizing the constraint violations in the action-value function dynamically by introducing the Lagrangian multiplier, instead of using a fixed penalty value in the reward [38, 39]. The multiplier is stochastically updated as a dual variable of the policy during the agent training based on the cost value function. However, the Lagrangian method-based Safe RL algorithm may not converge to the optimal solution because of the cost value function overestimation.

**Table 1**
Comparison of recently published ICES studies with the proposed model and optimization approach

| Reference | ICES modeling | | | | | | Optimization approach | | | |
|---|---|---|---|---|---|---|---|---|---|---|
| | Electric network | Gas network | Heat network | Realistic devices | Uncertainty | IDR | Non-convexity | Scalability | Privacy | Method |
| [9] | | | | | ✓ | ✓ | | | | Deterministic |
| [10] | | | | | ✓ | | | | | MPC |
| [11] | | | | | ✓ | | ✓ | | | GA |
| [12] | | | | | ✓ | | ✓ | | | NSGA-II |
| [13] | | | | | ✓ | | | | | Deterministic |
| [14] | ✓ | | ✓ | | ✓ | ✓ | ✓ | | | CDE |
| [15] | ✓ | | | | | | | | | Deterministic |
| [16] | ✓ | | ✓ | | ✓ | ✓ | | | | MPC |
| [17] | ✓ | ✓ | | | | | | | ✓ | ADMM |
| [18] | ✓ | | ✓ | | ✓ | ✓ | | | | RO |
| Proposed | ✓ | ✓ | ✓ | ✓ | ✓ | ✓ | ✓ | ✓ | ✓ | Safe RL |

*1.3 Summary of contributions*

In summary, three significant research gaps in the relevant literature have been identified to be addressed. 1) Most research overlooked the multi-network constraints in ICES. Several works accounting for network constraints have been done, but only two kinds of networks have been considered instead of multiple energies, including power, gas, and heat. However, the coordination of three types of integrated energy and their network constraints remains unrevealed but are crucial in improving energy efficiency. However, network constraints violations can compromise the economic value of the entire community and the security and stability of the network within ICES. 2) The ICES is not comprehensively modeled in previous research, which differs from the actual ICES operation and results in misleading analysis. Several vital models should be contained at once, including realistic devices, renewable uncertainty, and IDR of MEUs. These models describe the critical characteristics of flexibility and uncertainty on the demand side, and the interaction among them hugely affects the operational logic of ICES, which is unfortunately neglected by previous research. 3) Current methods cannot satisfy the requirements of ICES operational optimization in terms of solving non-convexity, meriting scalability, and protecting privacy. Approaches like mathematical programming not only require considerable work in convexification and linearization but also need complete information for an accurate solution, which breaches the privacy protection in ICES. The conventional RL algorithm, however, performs poorly in multi-network constrained optimization due to its unawareness of constraints in optimization logistics. 4) Existing Safe RL algorithms have multiple drawbacks, including difficulty in determining penalties for the direct penalization method, over-conservative policies for the Lyapunov method, and Q-value overestimation for the Lagrangian method. These problems result in an unfair tradeoff between the reward and cost during training, leading to severe constraint violations or conservative policies.

To our knowledge, the research on comprehensive techno–economic modeling and operational optimization in ICES, which considers multi-network constraints, is very limited. Analyzing previous literature, the reviewed studies are categorized in Table 1. As can be observed, this paper presents a novel Muti-Network Constrained ICES (MNC-ICES) model that considers network constraints of integrated energy, including electricity, natural gas, and heat. In the proposed model, the ICES operator (ICESO) secures the safe operation of the MNC-ICES considering non-convex energy devices, renewable uncertainties, and IDR of MEUs. A constrained optimization problem is formulated to denote the operation problem in the proposed MNC-ICES model and then transformed into a Constrained Markov decision process (C-MDP) for the application of RL approaches. Compared to existing software programs for regional IES operation, this research highlights the operation safety regarding detailed multi-network constraints and detailed energy device models. Moreover, a state-of-the-art (SOTA) Safe RL algorithm, namely primal-dual twin delayed deep deterministic policy gradient (PD-TD3), is developed based on the Lagrangian Safe RL method to optimize the scheduling and pricing strategies in MNC-ICES. The proposed algorithm shows the great potential for Safe RL to become a useful energy management tool in modern ICES regarding operational safety with multi-network.

The contributions of this paper are as follows:

1) A novel MNC-ICES model is proposed to interpret the concept of ICES. The proposed model accounts for the constraints of multi-network, which captures the physical characteristics of energy flow and imposes security operational constraints for the distribution level energy transmissions. Energy devices are modeled in high fidelity to describe the realistic physical operating attributes in practice. Additionally, the renewable uncertainty and integrated demand elasticity are considered to describe the novel characteristics of modern distribute-level energy systems. Overall, the proposed model can be implemented as a basis for practical network-constrained community operation tools.

2) A C-MDP is formulated from the constrained operational optimization problem in MNC-ICES with multi-energy integration. Constraints on voltage in the power network, gas flow, gas pressure and gas injection in the gas network, pipeline flow, and nodal flow in the district heat network are considered security constraints and imposed safety requirements, being modelled as the cost term in a tuple of C-MDP.

3) A novel Safe RL algorithm, namely PD-TD3, is proposed to solve the C-MDP and the constrained operational optimization problem in MNC-ICES. The PD-TD3 algorithm using double networks reduces the over-estimation problem of the action value for both the reward and cost, and the delayed update stabilizes the training process of policy and its dual variable. With such an accurate estimation of Q values, the proposed algorithm converges to the optimal solution that balances the maximal profits and the lowest constraint violation. In addition, the training processes of the policy and its dual variable are stabilized by delayed updates, which contributes to the training efficiency and helps to converge to the global optimal.

The remaining paper is organized as follows. The mathematical models of MNC-ICES, including integrated networks, energy devices, and MEUs, are presented in Section 2. The constrained operational optimization problem and the corresponding C-MDP are formulated in Section 3. The novel Safe RL algorithm is proposed in Section 4 to solve the C-MDP. Finally, several scenarios are simulated to verify the algorithm performance and analyze the simulation result in Section 5. The whole paper is concluded in Section 6.

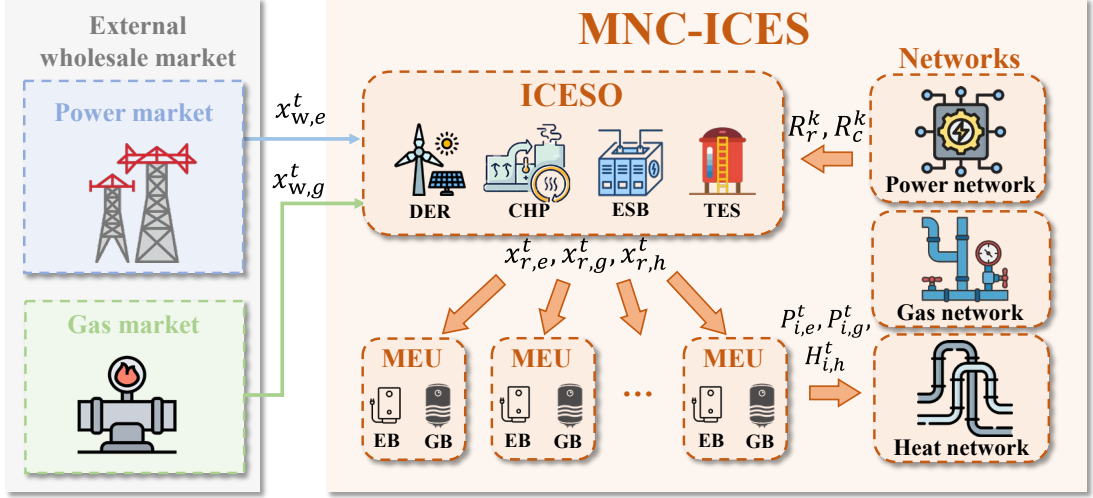

**Fig. 1.** Illustration of the proposed multi-network constrained integrated community energy system model

## 2. System modeling

This section proposes an MNC-ICES model, including various types of energy sectors and corresponding network models. Specifically, the MNC-ICES model, as depicted in Fig. 1, operates as a localized integrated energy system catering to MEUs on the demand side. The proposed model consists of 1) two types of DERs, WT and PV; 2) two types of energy storage systems, EBS and TES; 3) CHP as a power generation unit, as well as 4) MEU consisting of electric boiler (EB), gas boiler (GB), and energy demand for power and heat. More importantly, the modeling of physical integrated energy networks for electricity, natural gas, and heat within the MNC-ICES model is presented. These networks are foundational components and are vital for the efficient transmission and distribution of energy resources. To this end, multi-network constraints are proposed to govern the behavior of each network, complying with physical constraints in real-world operation. The cooling system (including CHP cooling generation, cooling network, and cooling load) is omitted for simplicity, since its similar operational characteristics to the heating system. The loads for MEUs are consequently modelled in terms of EB and GB, which is a simplified model but sufficient to reflect the basic consumption behavior in the regime of ICES operation.

The MNC-ICES model is assumed to encompass a singular operator, i.e., ICESO, scheduling energy devices and conducting energy transactions. The ICESO should manage the energy schedules of energy devices and determine the energy prices for MEUs to maximize the total profits without violating the network constraints. Therefore, the ICESO needs to schedule the energy devices dynamically for local energy conversion and price-integrated energy to mobilize the IDR resources of MEUs. In contrast, MEUs adjust energy consumptions due to IDR oriented from energy flexibilities. The whole period of operation and transaction can be divided into 24 intervals ($t = \{1, 2, \ldots, T\}$), and $N$ MEUs are represented by $i = \{1, 2, \ldots, N\}$. In each step, the ICESO should read the wholesale prices information, observe the local information on energy devices, and evaluate the state of charge of energy storage systems of TES and EBS before scheduling. Then, the energy prices for MEUs need to be set, the operation status of energy devices needs to be scheduled, and the TES and EBS need to be charged or discharged at each time interval. The detailed models of MNC-ICES are presented as follows.

### 2.1 Electricity distribution network

In the distribution of electricity networks, the prevailing topology is often radial, and it lends itself well to representation as a tree graph. In this representation, the root point corresponds to the connection with the transmission network. The distribution network can thus be visualized as an interconnected web of nodes and transmission lines, embodying the essential structure of a tree graph.

Let $n \in N_e$ denote the set of nodes within the distribution network, and $(n, m) \in P_e$ represent the set of transmission lines governing the interconnection of these nodes. Embracing the paradigm of radial distribution electricity networks, this network configuration captures the hierarchical nature of power flow from the root point, linked to the transmission network, branching out to various nodes within the distribution system. To govern and constrain the dynamics of real power, reactive power, and voltage within the radial distribution network, the linearized DistFlow approach is adopted [42]. The ensuing sections delve into the specifics of how linearized DistFlow constraints shape and guide the real power, reactive power, and voltage considerations in the context of distribution system operation.

$$P_1^t = \sum_{\forall n} x_n^t, t \in T \tag{1-a}$$

$$P_{n+1}^t = P_n^t - p_{n+1}^t, \forall n \in N_e, t \in T \tag{1-b}$$

$$Q_{n+1}^t = Q_n^t - q_{n+1}^t, \forall n \in N_e, t \in T \tag{1-c}$$

$$V_{n+1}^t = V_n^t - (b_n^1 P_n^t + b_n^2 Q_n^t), \forall n \in N_e, t \in T \tag{1-d}$$

$$\underline{V_n} < V_n^t < \overline{V}_n, \forall n \in N_e, t \in T \tag{1-e}$$

$$0 \le p_i^t \le \overline{p}_i^t, \forall i \in I, t \in T \tag{1-f}$$

$$0 \le q_i^t \le \overline{q}_i^t, \forall i \in I, t \in T \tag{1-g}$$

$$0 \le P_{nm}^t \le \overline{P}_{nm}, \forall n, m \in N_e, \forall (n, m) \in P_e, t \in T \tag{1-h}$$

$$0 \le Q_{nm}^t \le \overline{Q}_{nm}, \forall n, m \in N_e, \forall (n, m) \in P_e, t \in T \tag{1-i}$$

In (1) $P_{nm}^t$ and $Q_{nm}^t$ indicate the real power and reactive power flow from bus $n$ to node $m$ at time $t$. $V_n^t$ is the voltage magnitude at the bus $n$ at time $t$. $p_n^t$ and $q_n^t$ are the real and reactive power exchange at bus $n$. $b_n^1$ and $b_n^2$ are the resistance and reactance between the bus $n$ and $n + 1$. $\overline{V}_n / \underline{V}_n$ are upper/lower bound for voltages of each bus. $\overline{P}_{nm} / \underline{P}_{xm}$ and $\overline{Q}_{nm} / \underline{Q}_{nm}$ denote the upper/lower limits for active and reactive power of the transmission line between bus $n$ and bus $m$.

## 2.2 Natural gas distribution network

The natural gas network, renowned for its intricate network of pipelines enabling bidirectional gas flow, constitutes a critical infrastructure for the dissemination of energy resources. Traditionally, the directionality of gas flow is contingent upon the interplay of gas pressure differentials and injections at discrete nodes. However, this paper limits its scope to the dynamics of unidirectional gas flow within this network. This assumption is made upon operational constraints whereby consumers exclusively draw upon gas resources, with the absence of gas production and storage facilities.

Within this defined framework, let $n \in N_g$ denote the set of nodes, and $(n, m) \in P_g$ represent the set of gas pipelines intricately threading through the natural gas network. To model the dynamics of unidirectional gas flow, the study employs the Weymouth equation [43, 44]. The whole network constraints for natural gas networks are given as (12).

$$gf_{mn}^t = \text{sgn}(Pr_m^t, Pr_n^t) \, C_{mn} \sqrt{|(Pr_m^t)^2 - (Pr_n^t)^2|}, \forall (n, m) \in P_g, \forall t \in T \tag{2-a}$$

$$-\overline{gf}_{mn} \le gf_{mn}^t \le \overline{gf}_{mn}, \forall (n, m) \in P_g \tag{2-b}$$

$$G_n^t = -\sum_{m \in N_g} gf_{mn}^t, \forall (n, m) \in P_g, \forall t \in T \tag{2-c}$$

$$\underline{Pr}_n \le Pr_n^t \le \overline{Pr}_n, \forall n \in N_g, \forall t \in T \tag{2-d}$$

$$0 \le G_n^t \le \overline{G}_n, \forall n \in N_g, \forall t \in T \tag{2-e}$$

In (2), $gf_{mn}^t$ is the gas flow in the pipeline from node $m$ to node $n$. $Pr_n^t$ is the gas pressure of the node $n$. $G_n^t$ is the gas consumption in the node n. $C_{mn}$ is the line pack constant of gas pipeline $mn$. $\text{Sgn}(\cdot)$ is the signal function to determine the direction of the gas flow. Equations (2a)–(2c) show the constraints for nodal natural gas flow balance with the setting of $P_{REFg,t} = P_{n,max}$. In (2b), $\overline{gf}_{mn}$ is the limitations for the gas flow in the network. Equations (2d)–(2e) limit the nodal pressure and gas sources within its threshold, where $\overline{Pr}_n$ and $\underline{Pr}_n$ are the upper and lower bounds of gas pressure at node $n$, $\overline{G}_n$ is the limitation for gas consumption in node $n$. It is worth noting that (2a) is a non-convex equation constraint in an optimization problem, being hard to tackle by using a mathematical programming approach.

## 2.3 District heating network

Heat networks are vital for transmitting thermal energy through hot water via water pipelines, which are conventionally comprised of supply and return pipelines. The generation of heat energy, typically by CHP systems within the MNC-ICES model, initiates the flow of water in the supply pipelines to consumers at each node. After the consumer utilizes the heat energy, the water, now cooled, is directed back to the CHP through return pipelines. This unidirectional water flow mirrors the direction of heat flow. Notably, the temperature and pressure of water decrease along the heat transmission direction, indicating both heat loss during transmission and the propulsive force for water flow. The heat flow is roughly described by Fig. 2.

In this paper, Variable Flow Temperature Constant (VFTC) method is employed to model the heating network [45]. The temperature at the supply and return sides of each node is considered constant over time. During heat transmission, a fixed proportion of heat injected into a pipeline is lost as the water progresses to the next node. Denoting nodes as $n \in N_h$ and direct supply and return pipelines as $(n, m) \in S_n^+$ and $(n, m) \in S_n^-$, respectively, the heat network model is formulated as follows.

$$\sum_{(n,m) \in S_n^+} M_{nm}^t - \sum_{(n',m') \in S_n^-} M_{n'm'}^t = M_n^t, \forall (n,m) \in S_n^-, \forall (n',m') \in S_n^+, n \in N_h, \forall t \in T \qquad (3-a)$$

$$\sum_{i \in I_n} H_i^t = -c_f M_n^t (T_n^S - T_n^R), n \in N_h, \forall t \in T \qquad (3-b)$$

$$\underline{M_n^N} \leqslant M_n^t \leqslant \overline{M}_n^N, n \in N_h, \forall t \in T \qquad (3-c)$$

$$0 \leqslant M_{nm}^t \leqslant \overline{M}_{nm}^S, \forall (n,m) \in (S_n^- \cup S_n^+), \forall t \in T \qquad (3-d)$$

In (3), $M_{nm}^t$ represents the pipeline heating flow, $M_n^t$ denotes nodal heating flow, and $H_i^t$ signifies the nodal power injection of a consumer. $c_f$ denotes the heat capacity of water, while $T_n^S$ and $T_n^R$ indicate temperatures of node $n$ in the supply and return networks, respectively. $\overline{M}_n^N$ and $\underline{M}_n^N$ represent the upper and lower bounds of nodal flow. The heat flow $\overline{M}_{nm}^S$ in the pipeline $(n,m)$ is positive if the direction aligns with water flow and negative otherwise. In the proposed model, (3a) is the equality constraints for nodal flow, while (3-b) describes the nodal power injection given the nodal flow. Equations (3c) and (3d) impose inequality constraints on nodal flow and pipeline flow. Importantly, the constraints reveal bidirectional nodal flow and unidirectional water/heat flow within the pipeline.

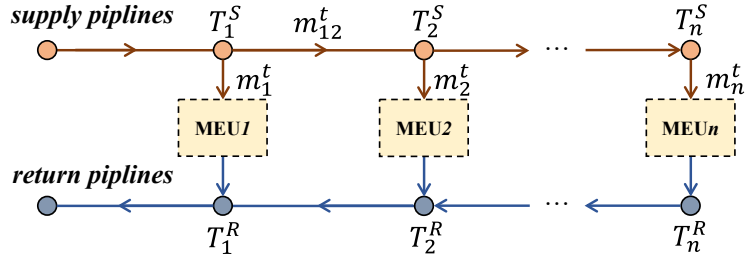

**Fig. 2.** Representation of district heating network

### 2.4 Energy devices modeling

#### 2.4.1 Combined heat and power (CHP)

CHP, a single-input multi-output energy converter, assumes a crucial part of the MNC-ICES model due to its high energy conversion efficiency from natural gas to electricity and heat [46, 47]. CHP is characterized by two constant energy conversion efficiencies for electricity and heat. The detailed operation model of CHP, depicted by a non-convex feasible operation region (FOR) enclosed by the boundary curve ABCDEFG, is adopted and shown in Fig. 3. $P_{CHP}^t$, $H_{CHP}^t$ are generated power and heat for the CHP in time slot t. The FOR of the CHP is divided into two convex sections and is represented as equations in Appendix.

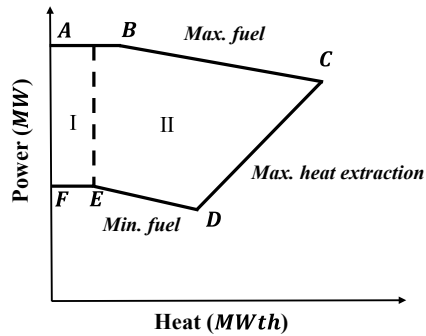

**Fig. 3.** Feasible operation region (FOR) of CHP units

#### 2.4.2 Distributed energy resources (DER)

The power output of DER, denoted as $P_{DER}^t$, is defined in equation (4) by incorporating the power generation of PV and WT. The power generation function for DER accounts for power output uncertainty, modeled by probabilistic distribution functions for PV and WT, respectively.

$$P_{DER}^t = P_{PV}^t + P_{WT}^t \qquad (4)$$

As variable renewable energy (VRE), wind power inherently carries high uncertainty. The wind speed ($\omega$), directly influencing power output, is predicted with an unavoidable error $\Delta\omega$, which is modelled by a Weibull PDF [48]. The power output $P_{WT}^t$ of WT is positive if and only if the wind speed exceeds the starting speed ($\omega_{in}^c$); otherwise, $P_{WT}^t$ is always zero. The upper limit for WT power is $P_{WT.rated}^t$ when $\omega_{rated}^c \leq \omega \leq \omega_{out}^c$. If the wind speed surpasses the cutout speed $\omega_{out}^c$, WT will be cut out, resulting in $P_{WT} = 0$. Additionally, the Weibull PDF is employed to estimate the uncertainty parameter due to wind speed prediction errors.

For photovoltaic power generation, this paper introduces the prediction error $\Delta I$ of PV in (8-a). PV generates electricity by converting solar radiation energy, and power generation is directly related to solar irradiance in (8-b). The Beta PDF is employed to estimate uncertain parameters with minimal error. The detailed expressions of the DER power generation and error sampling are shown in Appendix.

### 2.4.3 Energy Storage Systems (ESS)

ESS contains the EBS and TES for the energy storage of power and thermal energy, respectively. The EBS functions as a charge-dischargeable battery with varying efficiency [49]. The operational strategy of EBS is modeled at a granularity of one hour, i.e., one time interval. Charging and discharging operations are consolidated into a single activity within one time slot [50, 51]. In the model of EBS, $E_{EBS}^t$ is the battery capacity at time interval $t$. $\beta$ and $\eta_{EBS.c}$ are predetermined parameters representing the loss factor and charging efficiency, respectively. $P_{EBS.c}^t$ and $P_{EBS.d}^t$ represents the charging power and discharging power at time step $t$, respectively. $S_{EBS.c}^t$ and $S_{EBS.d}^t$ represent the charging state and discharging state at time step $t$, respectively. $P_{EBS.c.max}$ and $P_{EBS.d.max}$ are the maximum charging and discharging power, respectively. $E_{EBS.min}$ and $E_{EBS.max}$ represent the upper and lower limits of battery capacity, respectively. TES has a similar model to EBS. Please refer to Appendix for the detailed description.

### 2.5 Multi-energy user (MEU) modeling

As rational integrated energy consumers, MEUs engage in the procurement of energy to fulfill their energy demands for both power and heat [51]. MEUs are conceptualized to possess elastic electricity consumption appliances and energy conversion devices, including EB and GB, which enable them to adjust energy consumption dynamically across various time periods and multiple energies. Moreover, $E_{MEU}$ and $H_{MEU}$ are conventionally formulated as quadratic functions to encapsulate the utility derived from consumption for power and heat, respectively. This quadratic representation serves as an effective means to capture the nuanced relationship between power consumption and utility over time. As a result, MEUs can adjust their integrated energy consumption flexibly according to the inherent demand and dynamic energy prices, realizing the IDR to the community.

## 3. Problem formulation

This section presents the multi-networks constrained operational optimization problem for the ICESO and reformulates it into a corresponding C-MDP for the implementation of Safe RL algorithm. Specifically, the cost term in the C-MDP is denoted by the network constraint violations. By solving this C-MDP, the ICESO can maximize its reward with the tolerated constraint violation.

### 3.1 Objective function and constraints

The profit of ICESO is mainly the difference between the revenue for selling energy to MEUs and the cost of energy purchasing, as well as the imbalance penalty. The corresponding objective function is presented in (5).

$$\max_{\varphi} U_{IESP} = \sum_{t=1}^{T} \left\{ \begin{array}{l} \left( \underbrace{x_{r.e}^t \sum_{i=1}^{N} P_{i.e}^t + x_{r.g}^t \sum_{i=1}^{N} P_{i.g}^t + x_{r.h}^t \sum_{i=1}^{N} P_{i.h}^t}_{Revenue\ for\ selling\ energy} \right) - \\ \underbrace{\left( x_{w.e}^t P_{w.e}^t + x_{w.g}^t P_{w.g}^t \right)}_{Cost\ for\ energy\ pruchase} - \underbrace{\left( \delta_e^t P_{imb.e}^t + \delta_g^t P_{imb.g}^t + \delta_h^t H_{imb}^t \right)}_{Cost\ for\ energy\ balance} \end{array} \right\} \quad (5-a)$$

$$s.t. \ \forall t \in T$$

$$(1) - (3)$$

$$P_{w.e}^t + P_{DER}^t + \sum_{n \in N_e} P_{CHP.n.e}^t + P_{EBS.d}^t - P_{EBS.c}^t + P_{imb.e}^t = \sum_{i \in I_n} P_{i.e}^t \quad (5-b)$$

$$\sum_{n \in N_h} H_{CHP.n}^t + H_{TES.d}^t - H_{EBS.c}^t + H_{imb}^t = \sum_{i \in I_n} H_i^t \quad (5-c)$$

$$P_{w.g}^t + P_{imb.g}^t = \sum_{i \in I_n} P_{i.g}^t + \sum_{n \in N_g} P_{CHP.g}^t \quad (5-d)$$

$$x_{min.e}^t \leq x_{r.e}^t \leq x_{max.e}^t \quad (5-e)$$

$$x_{min.g}^t \leq x_{r.g}^t \leq x_{max.g}^t \quad (5-f)$$

$$x_{min.h}^t \leq x_{r.h}^t \leq x_{max.h}^t \quad (5-g)$$

In (5), $\varphi = \{x_{r.e}^t, x_{r.g}^t, x_{r.h}^t, P_{CHP.e}^t, P_{EBS.d}^t, P_{EBS.c}^t, H_{CHP}^t, H_{TES.d}^t, H_{TES.c}^t\}$ is the set of decision variables, and several decision variables are omitted due to the energy balance among several variables. The objective function in (5-a) constitutes three parts, revenue for selling energy, cost for energy purchase, and cost for energy balance, where $P_{imb.e}^t$, $H_{imb}^t$, $P_{imb.g}^t$ are the imbalanced electricity, heat and natural gas for ICESO. Penalty indexes $\delta_e^t$, $\delta_g^t$, $\delta_h^t$ are preset parameters to penalize the energy imbalance and determined based on energy prices. Also, the objective is constrained by (1)-(3) and (5-b)-(5-g). Equality constraints (5-b)-(5-d) indicate the integrated energy balance. (5e)-(5g) are inequality constraints for the retail energy prices, where $x_{max.e}^t$, $x_{min.e}^t$, $x_{max.g}^t$, $x_{min.g}^t$, $x_{max.h}^t$, $x_{min.h}^t$ are preset parameters indicating the upper bounds and lower bounds for power, natural gas and heat, respectively.

The energy balance constraints are actually relaxed by introducing the penalty terms $\delta$. However, network constraints are not directly relaxed to the objective function, as penalties for network constraint violations are hard to determine. Specifically, compared with energy imbalance that only decreases the profits from the economic perspective, the violation of network constraints is more serious and may affect the safe operation of the ICES. Moreover, determining penalties for network constraint violation to realize a fair tradeoff between improving profits and reducing violations is not straightforward. Therefore, it is assumed that the ICESO aims to guarantee safe operation rather than uplift the economic revenue. Consequently, the network-constrained operational optimization problem is formulated to C-MDP as follows.

### 3.2 Markov decision process (MDP)

To optimize the decision-making process of ICESO, a MDP is leveraged to describe the integrated energy transactions and then a DRL algorithm is used to solve it. This approach treats the ICESO as an intelligent agent that makes decisions based on the environmental observation of wholesale market prices (both electricity and gas), and power generation of DER. The objective is to improve the pricing decisions by maximizing the accumulated return, using a well-defined reward function in (14-a). The MDP can be denoted by $< S, A, R, P, \mu, \gamma >$. $S$ is the set of states. $S = \{x_{w.e}^t, x_{w.g}^t, P_{WT.predict}^t, P_{PV.predict}^t, E_{EBS}^t, E_{TES}^t\}$, encompassing electricity market price, natural gas market price, forecast power generation of WT and PV, state of charge of EBS and TES. $A$ is the set of actions. $A = \{x_{r.e}^t, x_{r.g}^t, x_{r.h}^t, P_{CHP}^t, H_{CHP}^t, P_{EBS.c}^t, P_{EBS.d}^t, H_{TES.c}^t, H_{TES.d}^t\}$ represents the available actions as the decision variables in (14-a). $R: S \times A \times S \mapsto \mathbb{R}$ is the reward function, which quantifies the action's performance and is presented by the objective function. $P: S \times A \times S \mapsto [0, 1]$ is the transition probability function. The state transition function is not considered due to the assumption of uncoupled state across time periods. $\mu: S \mapsto P(A)$ represents the policy of the agent, mapping from states to a probability distribution over actions [52]. $\gamma \in [0, 1]$ is the discount factor to discount the future reward.

The discounted accumulative reward under policy $\mu$ is denoted as (6). In (6), $\tau = (s_0, a_0, s_1, a_1 \cdots)$ is a trajectory of the agent with a series of actions, and $\tau \sim \pi$ indicate trajectories distribution under policy. To conclude, the aim of MDP shown in (16) is to find the optimal policy $\mu^*$ that can maximize the discounted accumulative reward $R(\mu)$.

$$R(\mu) = \mathbb{E}_{\tau \sim \pi} \left[ \sum_{t=0}^{\infty} \gamma^t R(s_t, a_t, s_{t+1}) \right] \tag{6}$$

$$\mu^* = \arg \max_{\mu} R(\mu) \tag{7}$$

### 3.3 Constrained-Markov decision process (C-MDP)

To maintain the energy flow complying with the network constraints, a cost function is proposed for the C-MDP, indicating the violation of network constraints. The C-MDP can be denoted as $< S, A, R, C, P, \mu, \gamma >$, which is an ordinary MDP augmented by cost function $C(s, a)$. The cost function is denoted by $C: S \times A \times S \mapsto \mathbb{R}$, mapping from transition tuples to cost. The explicit expression of the cost function is given in (8). It comprises the standardized constraint violation of cost in three kinds of network constraints. As the transmission capacity is always designed to be large enough to carry the real and reactive power in the electric distribution network, only voltage constraints are considered in the following cost function.

$$C = \left\{ \begin{array}{l} \underbrace{\sum_{n \in N_e, \forall (n,m) \in P_e} \left[ \left[ \frac{V_{n,t} - \overline{V}_n}{\overline{V}_n} \right]^+ + \left[ \frac{\underline{V}_n - V_{n,t}}{\underline{V}_n} \right]^+ \right]}_{\text{Cost for constraints violation in electricity network}} + \\ \underbrace{\sum_{n \in N_g, \forall (n,m) \in P_g} \left[ \left[ \frac{|gf_{k,mn}| - \overline{gf}_{mn}}{\overline{gf}_{mn}} \right]^+ + \left[ \frac{Pr_{k,n} - \overline{Pr}_n}{\overline{Pr}_n} \right]^+ + \left[ \frac{\underline{Pr}_n - Pr_{k,n}}{\underline{Pr}_n} \right]^+ + \left[ \frac{G_{k,n} - \overline{G}_n}{\overline{G}_n} \right]^+ \right]}_{\text{Cost for constraints violation in gas network}} + \\ \underbrace{\sum_{n \in N_h, \forall (n,m) \in P_h} \left[ \left[ \frac{M_n^t - \overline{M}_n^N}{\overline{M}_n^N} \right]^+ + \left[ \frac{\underline{M}_n^N - M_n^t}{\underline{M}_n^N} \right]^+ + \left[ \frac{M_{nm}^t - M_{nm}^S}{\overline{M}_{nm}^S} \right]^+ \right]}_{\text{Cost for constraints violation in heat network}} \end{array} \right\} \tag{8}$$

In (8), $[x]^+ = \max\{0, x\}$ is the projection function. The cost function is constituting costs for constraint violation in the electricity network, gas network, and heat network. To limit the constraint violation, the constraint for the cost function is proposed as (9), where $d$ is the upper bound of the cost function.

$$C(\mu) \leq d \qquad (9)$$

The long-term discounted cost under policy $\mu$ is similarly defined as $C(\mu) = E_{\tau \sim \mu}[\sum_{t \in T} \gamma_t \, C(s_t, a_t, s_{t+1})]$, and the corresponding limit is $d$. In the C-MDP, the goal is to select a policy $\mu$ that maximizes the long-term reward $R(\pi)$ while satisfying the constraints on the long-term costs.

$$\mu^* = \arg\max_\mu R(\mu) \qquad (10)$$
$$s.t. \,(9)$$

## 4. Primal-dual deep deterministic policy gradient (PD-TD3) algorithm

In this section, a PD-TD3 algorithm is developed to solve the proposed C-MDP and learn the optimal operational strategy for ICESO. Specifically, the proposed C-MDP is formulated into a Lagrangian function, which is then converted to an unconstrained min-max problem and thus applicable to the solution of the iterative primal-dual TD3 algorithm. The PD-TD3 algorithm then solves the primal-dual problem by using the gradient descent to iteratively update the policy and Lagrangian multiplier.

### 4.1 Primal-Dual TD3 algorithm

The challenges of optimal operation of MNC-ICES model mainly come from the non-linear integrated network constraints. Conventional deep reinforcement learning algorithms do not directly consider these constraints in the learning process [52]. Moreover, traditional DRL algorithms still face the problem of overestimation of Q-value and cost value in C-MDP and instability during the training process. To overcome these drawbacks, the proposed PD-TD3 algorithm is able to address the challenges of constrained optimal operation problem in MNC-ICES model by solving the C-MDP, and mitigating the issue of value overestimation and training instability.

As a RL method, the key of the primal-dual algorithm is to augment constraints on the expected rewards, such that the training of the RL agent converges to the optimal constraints-satisfying policies. Therefore, the objective of primal-dual TD3 for cost minimization can be generally written as (20), where $\mathcal{L}(\mu, \lambda)$ is the augmented objective action-value function, $\lambda$ denotes the multipliers of constraints.

$$\mathcal{L}(\mu, \lambda) = R(\mu) - \sum_i \lambda(C(\mu) - d) \qquad (11)$$

$$(\mu^*, \lambda^*) = \arg\min_{\lambda > 0} \max_\mu \mathcal{L}(\mu, \lambda) \qquad (12)$$

In (11), $R(\mu)$ and $C(\mu)$ represent the reward and the cost for constraint violation of a DRL agent. For constrained optimal operation of MNC-ICES model, the reward and constraint violation can be the total profits of the ICESO and violation of physical constraints of integrated distribution networks, respectively. To solve the unconstrained minimax problem (12), the iterative primal-dual method is used as a canonical approach where in each iteration. In each iteration, the primal policy $\pi$ and the dual variable $\lambda$ are updated in turn. The primal-dual update procedures at iteration $k$ are as follows:

$$\theta_{k+1} = \theta_k + \alpha_k \nabla_\theta (\mathcal{L}(\mu(\theta), \lambda_k))|_{\theta = \theta_k} \qquad (13)$$
$$\lambda_{k+1} = f_k(\lambda_k, \mu(\theta)) \qquad (14)$$

In the proposed PD-TD3 algorithm, the primal variable, i.e., policy parameters, is updated by policy gradient, which is specified later. The dual variable, i.e., the Lagrangian multiplier, is updated by (15). In (15), $\beta_k$ is the step size of the multiplier update. $[x]^+ = \max\{0, x\}$ is the projection onto the dual space $\lambda^k > 0$.

$$\lambda_{k+1} = [\lambda_k + \beta_k(C(\mu_k) - d)]^+ \qquad (15)$$

### 4.2 Algorithm Design for Primal-Dual TD3

The proposed PD-TD3 algorithm is an off-policy DRL algorithm, enabling offline training of strategies in optimization problems and using DNN approximate action-value functions. The overall framework of the PD-TD3 is summarized in Algorithm 1. As a DRL algorithm based on the actor-critic framework, the PD-TD3 adopts DNNs to approximate the value functions and policy functions of the C-MDP, which denotes the critic and actor, respectively. To estimate both the reward and cost in the C-MDP, PD-TD3 employs two kinds of critic networks, namely the reward critic network and the cost critic network. Additionally, as PD-TD3 uses the trick of double networks, each type of critic consists of two online Q networks and their target networks, mitigating the issue of Q-value overestimation observed in other value-based RL algorithms. Also, the target networks of the critic are delayed copies of the online network, which is supposed to mitigate the instability of the training process. Therefore, three sets of neural networks are employed: (1) two reward critic Q-networks $Q_{R1}(s, a | \theta_{R1}^Q)$, $Q_{R2}(s, a | \theta_{R2}^Q)$ and their target network $Q_{R1}'(s, a | \theta_{R1}^{Q'})$, $Q'_{R2}\left(s,a|\theta^{Q'}_{R2}\right)$, (2) two cost critic Q-networks $Q_{C1}\left(s,a|\theta^Q_{C1}\right)$, $Q_{C2}\left(s,a|\theta^Q_{C2}\right)$ and their target networks $Q'_{C1}\left(s,a|\theta^{Q'}_{C1}\right)$, $Q'_{C2}\left(s,a|\theta^{Q'}_{C2}\right)$, and (3) the actor policy network $\mu(s|\theta^\mu)$ and its target network $\mu'(s|\theta^{\mu'})$.

During the training process, the agent randomly samples transitions $(s_i, a_i, r_i, c_i, s_{i+1})$ from the experience replier buffer (ERB) to form a mini-batch $N$ for experience replay learning. Then, the target of the reward and cost critic networks are presented as (16) and (17), which are employed to update Q-functions.

$$y_i = r_i + \gamma \min_{j\in\{1,2\}} Q\left(s_{i+1}, \tilde{a}_{i+1}|\theta^{Q'}_{Rj}\right) \tag{16}$$

$$z_i = c_i + \gamma \min_{j\in\{1,2\}} Q\left(s_{i+1}, \tilde{a}_{i+1}|\theta^{Q'}_{Cj}\right) \tag{17}$$

In (16) and (17), $\tilde{a}_{t+1}$ is the clipped target action shown in (18). Here, target policy smoothing is employed by incorporating clipped Gaussian noise into the target action during the evaluation process. This technique promotes smoother and more stable policy updates, facilitating convergence and enhancing the quality of the learned policy.

$$\tilde{a}_{i+1} = \mu'(s|\theta^{\mu'}) + \tilde{\varepsilon}, \tilde{\varepsilon} \sim clip(\mathcal{N}(0,\tilde{\sigma}), -c, c) \tag{18}$$

Based on the target, the reward and cost critic networks are updated by minimizing the loss function, i.e., mean square error (MSE) between the value functions and their targets, proposed in (19) and (20), respectively.

$$L_R = \frac{1}{N}\sum_{i\in N}[y_i - Q_R(s_i, a_i|\theta^Q_R)]^2 \tag{19}$$

$$L_C = \frac{1}{N}\sum_{i\in N}[z_i - Q_C(s_i, a_i|\theta^Q_C)]^2 \tag{20}$$

To mitigate the training error caused by correlated samples, the primal variable, i.e., policy network, and dual variable, i.e., Lagrangian multiplier after a fixed number $e$ of iterations by using (21) and (22), which is the so-called "delayed" update. This delay in primal and dual variable updates reduces the correlation between successive updates and prevents rapid forgetting of previously learned policies. The policy is updated by one step of sampled gradient descent using (21). The Lagrangian multiplier is also updated by using sampled dual gradient in (31). Also, it should be noted that the dual variable updated in (22) uses the minimized estimated Q-value for cost, alleviating the overestimation of the Q value to ensure a proper update descent.

$$\nabla_{\theta^\mu}\mathcal{L}(\theta^\mu, \lambda) \approx \frac{1}{N}\sum_{i\in N}\nabla_{\theta^\mu}\left[Q_{R1}\left(s_i, \mu(s_i|\theta^\mu)|\theta^Q_{R1}\right) - \lambda Q_{C1}\left(s_i, \mu(s_i|\theta^\mu)|\theta^Q_{C1}\right)\right] \tag{21}$$

$$\nabla_\lambda \mathcal{L}(\theta^\mu, \lambda) = \frac{1}{N}\sum_{i\in N}\left[\min_{j\in\{1,2\}} Q\left(s_{i+1}, \tilde{a}_{i+1}|\theta^{Q'}_{Cj}\right) - d\right] \tag{22}$$

Additionally, all target networks of the actor and critic are updated by using the soft update presented in (23) and (24). It allows a small pace update in each iteration and ensures a gradual and stable convergence of the networks, where $\rho$ represents the soft update parameter.

$$\theta^{Q'} \leftarrow \rho\theta_Q + (1-\rho)\theta^{Q'} \tag{23}$$

$$\theta^{\mu'} \leftarrow \rho\theta_\mu + (1-\rho)\theta^{\mu'} \tag{24}$$

### 4.3 Discussion of Potential Limitations

Previous subsections address the Q-value overestimation problem in the typical RL algorithms. Considering the PD-TD3 is developed based on the conventional TD3 algorithm, it also inherits the following drawbacks: 1) The PD-TD3 algorithm is more complex than the TD3 algorithm and requires more computing resources by increasing numbers of hyperparameters. 2) The TD3-based algorithm is relatively sensitive to the selection of hyperparameters. However, as the ultimate goal is to deploy this well-trained algorithm to online dispatch, this could not be a serious problem.

If this algorithm is deployed to real-world ICES for online dispatch, an important assumption is that, the environment (state transition) of the simulation (test system) should be similar to the real-world ICES. Otherwise, the algorithm will generate unsafe decisions because it cannot be adaptive to the unknown environment. Potential solutions are twofold. First, a comprehensive modeling of system state transition using advanced deep learning methods is necessary. Second, the output of this RL algorithm should be corrected by real-time control algorithms, such as model predictive control.

---

**Algorithm 1** PD-TD3 algorithm

---

1:  Initialize policy parameters $\theta^\mu$, Q-function parameters $\theta^Q_{R1}$, $\theta^Q_{R2}$, $\theta^Q_{C1}$, $\theta^Q_{C2}$, and empty buffer $R$

2:  Initialize target networks $\theta^{\mu'} \leftarrow \theta^\mu$, $\theta^{Q'}_{R1} \leftarrow \theta^Q_{R1}$, $\theta^{Q'}_{R2} \leftarrow \theta^Q_{R2}$, $\theta^{Q'}_{C1} \leftarrow \theta^Q_{C1}$, $\theta^{Q'}_{C2} \leftarrow \theta^Q_{C2}$

3:  Initialize Lagrangian multiplier $\lambda$

4:  **repeat**

5:      Initialize a random process $N$ for action exploration

6:      Receive initial state $s_0$

7:  **for** each transaction time slot $t = 1, \ldots, T$ **do**
8:    Select action
9:    Execute action and observe
10:   Store transition in the reply buffer
11:   if s is terminal, reset environment state
12:   Randomly sample a bath of transitions from $N$
13:   Compute target actions using

$$\tilde{a}_{i+1} = \mu'(s|\theta^{\mu'}) + \tilde{\varepsilon}, \tilde{\varepsilon} \sim clip(\mathcal{N}(0, \tilde{\sigma}), -c, c)$$

14:   Compute target using

$$y_i = r_i + \gamma \min_{j \in \{1,2\}} Q\left(s_{i+1}, \tilde{a}_{i+1}|\theta_{R_j}^{Q'}\right)$$

$$z_i = c_i + \gamma \min_{j \in \{1,2\}} Q\left(s_{i+1}, \tilde{a}_{i+1}|\theta_{C_j}^{Q'}\right)$$

15:   Update Q-function by one step of gradient descent by minimizing

$$L_r = \frac{1}{N} \sum_{i \in N} [y_i - Q_r(s_i, a_i|\theta_R^Q)]^2$$

$$L_c = \frac{1}{N} \sum_{i \in N} [z_i - Q_c(s_i, a_i|\theta_C^Q)]^2$$

16:   **if** $k$ mod $e$ **then**
17:     Update policy by one step of gradient ascent using

$$\nabla_{\theta^\mu} \mathcal{L}(\theta^\mu, \lambda) \approx \frac{1}{N} \sum_{i \in N} \nabla_{\theta^\mu} [Q_{R1}(s_i, \mu(s_i|\theta^\mu)|\theta_{R1}^Q) - \lambda Q_{C1}(s_i, \mu(s_i|\theta^\mu)|\theta_{C1}^Q)]$$

18:     Update Lagrangian multiplier by one step of gradient ascent using

$$\nabla_\lambda \mathcal{L}(\theta^\mu, \lambda) = \frac{1}{N} \sum_{i \in N} \left[ \min_{j \in \{1,2\}} Q\left(s_{i+1}, \tilde{a}_{i+1}|\theta_{C_j}^{Q'}\right) - d \right]$$

19:     Update target networks with

$$\theta^{Q'} \leftarrow \rho \theta_Q + (1-\rho)\theta^{Q'}$$

$$\theta^{\mu'} \leftarrow \rho \theta_\mu + (1-\rho)\theta^{\mu'}$$

20:   **end if**
21:  **end for**
22: **until** convergence

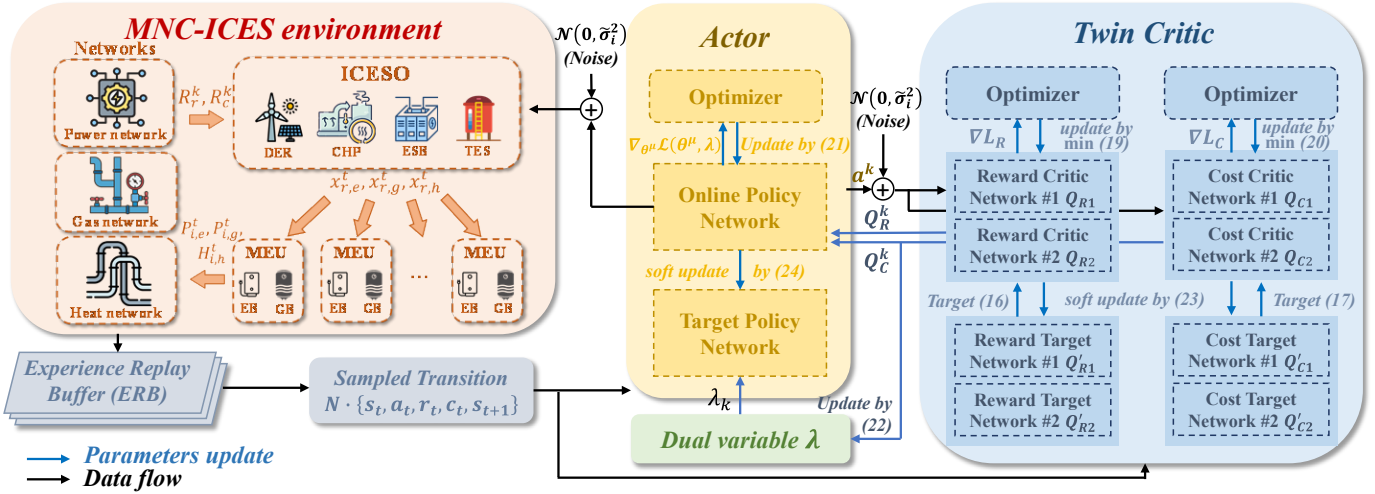

**Fig. 4.** Illustration of the proposed PD-TD3 algorithm

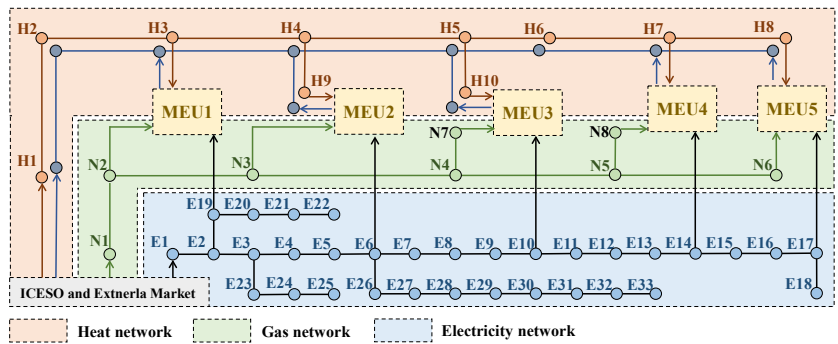

**Fig. 5.** Test system of integrated community energy system

**Table 2**
Neural network architectures settings

| Neutral Networks | Actor | Critic |
|---|---|---|
| No. of Hidden Layers | 3 | 2 |
| No. of Neurons | [128,32] | [128,32] |
| Activation Function | tanh | ReLU |
| Learning Rate | 4e-4 | 7e-4 |
| Soft update parameter | 1e-3 | 1e-3 |
| Delayed update frequency | 2 | 2 |
| Optimizer | Adam optimizer | Adam optimizer |

**Table 3**
Hyper-parameters settings

| Training parameters | Parameters |
|---|---|
| Replay Buffer Size | 1e+6 |
| Replay Start Size | 128 |
| Batch Size | 128 |
| Discount Factor | 0.99 |

## 5. Numerical results

### 5.1 Simulation setup

To validate the performance of the proposed PD-TD3 algorithm, a test system consisting of 5 MEUs is adopted and is shown in Fig. 5. As shown in the test system, a standard IEEE-33 bus electricity network, a heat network, and a natural gas network are considered to model the whole integrated energy network structure in MNC-ICES. It should be mentioned that simulation based on the test system is a generalized scenario but not a representation of a specific real-world application. The numerical results obtained from the proposed test system only serve as a demonstration of the proposed models and methods and a foundation for further application. To bridge the gap between generalized scenarios and real-world applications of the ICES models and Machine Learning methods, future work will involve applying the developed methods to actual energy systems models or more detailed case studies.

The key parameter settings for the test system are given as follows. The voltage constraints of the electricity network are set with an upper bound of $0.9$p. u. and lower bound of $1.1$p. u., and the other configuration data for the power network is taken from [53]. The natural gas network transmits the gas in the pipeline with an inside diameter of $0.3$m and an efficiency of 90%, operating at a temperature of $288.15°K$. The gas compressibility factor is set as $0.9 Pa^{-1}$. The allowable pressure for gas transmission is limited from $110 kPa$ to $100 kPa$, and the maximal gas flow rate is $400 m^3/h$. In the 10-node district heat system, the supply temperature at the most upstream node is $70°C$, and the return side temperature of the most downstream node is set to $30°C$. The temperature loss is assumed to be $0.1$K/m on the supply side, and $0.05$K/m in return side pipelines [50].

The hourly wholesale prices for electricity and natural gas are obtained using the real data of New England ISO. The constant natural gas price is set at $4.75£$/MMBtu, resulting in a natural gas price of approximately $16.2$ $£$/MW or $0.165$ $£$/m$^3$. Moreover, the power output of WT and PV are adapted from real-world data in [54]. The pricing ranges for electricity, natural gas, and heat in the MNC-ICES are set as $0-50£$/MW, $0-50£$/MW, and $0-40£$/MW, respectively.

The proposed DRL algorithm is implemented on the Pytorch [55]. The neural networks are configured with the settings shown in Table 2. The hyperparameters of the algorithm shown in Table 3 are selected based on empirical values and adjusted during the training process until the algorithm converges to the maximum profit. The quadratic programming problem for the comprehensive energy demand response of the MEUs is solved using commercial solver.

### 5.2 Training performance

This subsection aims to validate the convergence performance of the PD-TD3. Safe RL algorithms using both the Lagrangian method and the direct penalty method are employed as benchmark algorithms. The L-SAC [35] and S-DDPG [38] belong to the former, while typical TD3 with direct penalizing cost stands for the latter. The penalty index $\lambda$ is set as 1, 10, 100, and 1000 for a comprehensive comparison. In this context, each algorithm is trained 1000 episodes to learn the optimal strategy in pricing and scheduling in MNC-ICES, while each episode contains 24 steps, indicating 24 hours per day. Figs. 6-7 present the evolution of cumulative reward and cost for each episode, respectively. The corresponding values are also listed in Table 4 for a clearer demonstration. The allowable operating range of the cumulative cost is 0~10, and the upper bound of 10 is marked in black in Fig. 7.

As illustrated in Fig. 6, the cumulative reward of the four algorithms has a similar trend, which fluctuates a lot in the initial stage of training, since the algorithm randomly chooses actions to explore the environment. Similarly, the initial cost for constraint violation in Fig. 7 is relatively higher and fluctuates in the initial stage. With the learning process going on, the policy is continuously trained and improved, resulting in an increasing trend in reward and a decreasing trend in cost. In the comparison between the algorithms using the Lagrangian-based method and direct penalty method, it can be observed that typical TD3 algorithms with fixed penalty index usually have lower reward and higher cost. The reward and cost of TD3 decrease as $\lambda$ increases. Specifically, TD3 with $\lambda = 1000$ has the lowest reward among all algorithms even L-SAC, and the lowest cost among TD3 with all $\lambda$ settings. Moreover, its cost reaches around the allowable requirement but is still unqualified. This demonstrates a worse performance of the direct penalty-based Safe RL algorithms compared to Lagrangian-based Safe RL algorithms.

In the comparison within algorithms using the Lagrangian Safe RL method, it can be observed that the L-SAC converges with the lowest cumulative reward, which is nearly zero. This is thought as a local optimum and caused by the improper tradeoff between the reward and cost, indicating an over-conservative policy of L-SAC. However, the PD-TD3 shows a fast convergence to the highest reward among the three algorithms, and it can be observed in both Fig. 6 and Table 4 that the cumulative reward is about to converge around 200 episodes with a reward of almost 10000. This is driven by the delayed policy update that can update the policy without the training noise, training the policy networks effectively. In addition, the PD-TD3 deals with the physical constraints by directly adjusting the policy of the actor-network. On this account, the cost of PD-TD3 for an episode containing 24 operating hours is in the allowable range of 0~10 after 500 episodes. In comparison, the cost of S-DDPG is out of the allowable range during almost the whole training process, while the cost of L-SAC is about 0 and is thought of as over-conservative. This is owing to the double Q cost networks that assist in estimating the cost more precisely by eliminating the overestimation of cost and thus achieving a fair balancing of the tradeoff between the reward and cost, which has the highest reward and an allowable cost. Furthermore, it can be observed that the reward of PD-TD3 converges around 200 episodes, and the cost is operating in the allowable range after about 500 episodes. The convergence speed of PD-TD3 is similar to L-SAC but is much faster than S-DDPG twice. Also, the convergence process of the reward and cost in PD-TD3 demonstrates that the dual variable converges to optimal after the convergence of reward, since the dual variable is updated with delay in a small step to allow a high exploration in reward, avoiding getting stuck in a local optimum like L-SAC.

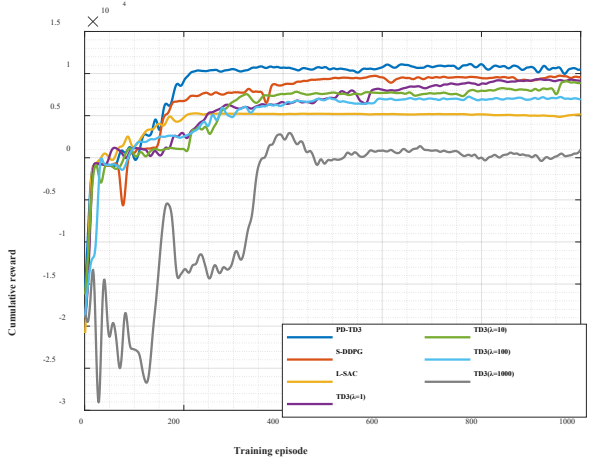

**Fig. 6.** The evolution of cumulative reward for different Safe RL algorithms.

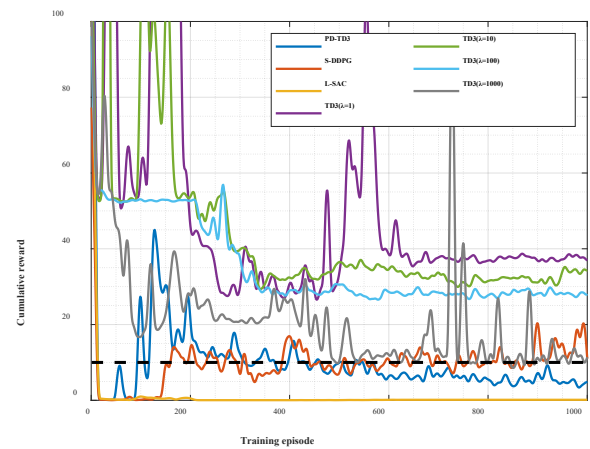

**Fig. 7.** The evolution of cumulative cost of constraint violation for different Safe RL algorithms.

**Table 4**
The cumulative values of reward and cost for constraint violation for different Safe RL algorithms.

| Algorithms | Evaluation | Episode | | | | | | |
|---|---|---|---|---|---|---|---|---|
| | | 1 | 100 | 200 | 400 | 600 | 800 | 1000 |
| PD-TD3 | Reward | -15402 | -87 | 9113 | 10703 | 11105 | 10513 | 10473 |
| | Cost | 196 | 27 | 21 | 12 | 6 | 5 | 5 |
| S-DDPG | Reward | -18964 | 556 | 6744 | 8665 | 9462 | 9596 | 9555 |
| | Cost | 77 | 1 | 14 | 17 | 11 | 10 | 11 |
| L-SAC | Reward | -20826 | 1282 | 4778 | 5179 | 5151 | 5145 | 5141 |
| | Cost | 101 | 1 | 0.5 | 0 | 0.1 | 0.1 | 0.1 |
| TD3 ($\lambda = 1$) | Reward | -18965 | 1151 | 2786 | 6593 | 8794 | 9162 | 7412 |
| | Cost | 170 | 60 | 46 | 30 | 38 | 37 | 37 |
| TD3 | Reward | -16229 | 335 | 1022 | 7562 | 7964 | 8922 | 6925 |

| (λ =10) | Cost | 126 | 105 | 54 | 32 | 35 | 32 | 34 |
|---|---|---|---|---|---|---|---|---|
| TD3 | Reward | -18967 | 1266 | 2530 | 6254 | 6932 | 6987 | 6614 |
| (λ =100) | Cost | 96 | 53 | 53 | 29 | 29 | 28 | 28 |
| TD3 | Reward | -18964 | -22773 | -13324 | 2363 | 7 | 1015 | 6614 |
| (λ =1000) | Cost | 8990 | 1830 | 2102 | 1993 | 2072 | 4210 | 4471 |

### 5.3 Generalization performance

To demonstrate the generalization performance of the proposed approach, two scenarios are simulated from the data set and analyzed in Figs. 11-12. Two scenarios are characterized as typical days for summer and winter with different demands and renewable generation. Despite the two scenarios having different energy production and consumption characteristics, there are some similarities during the operation. Firstly, during periods without lower demands for electric and heat power, the CHP unit is turned off due to its high operational cost; the demands are satisfied mainly by imported power and gas. Secondly, electricity prices show a similar trend to the homogenous demand and wholesale prices for electric power across scenarios. These show the generic strategy of the learned policy when facing similar conditions in ICES operation.

Nevertheless, the learned policies and energy resources show more differences across scenarios. As shown in Fig. 11, the CHP unit in winter day is turned on for about 15 hours on winter days (0:00-11:00 and 18:00-22:00) since profits caused by high demand for electricity and heat power across most hours can cover the operational cost of CHP. However, CHP only works one hour on summer day in Fig. 10 due to the lower demands and potential uneconomic operation. This results in heavy reliance on the external market on the summer day. Moreover, Fig. 9 shows lower prices for electric and heat power on winter day. This is a consequence of the power generation of the CHP unit and wind turbine, which has a lower cost than the external prices or zero generation cost. On the other hand, the flexibility of ESS is more efficiently realized on the summer day in Fig. 8. Although the export of power is not allowed in the proposed model, the larger dependence on the external market on the summer day provides more arbitrary chances for ESS.

Finally, it can be concluded that the proposed PD-TD3 algorithm is able to learn an effective policy for profit-maximization and safe operation in an MNC-ICES that can generalize to different scenarios. Furthermore, the proposed method investigates the flexibility potential of energy sources for two typical summer and winter days. More specifically, compared to the summer day, the ICES reports on CHP electric and heat power generation on winter day due to its less renewable power generation and higher demands. In contrast, summer day imports more electric power and natural gas from the external market, leading to higher energy prices for MEUs.

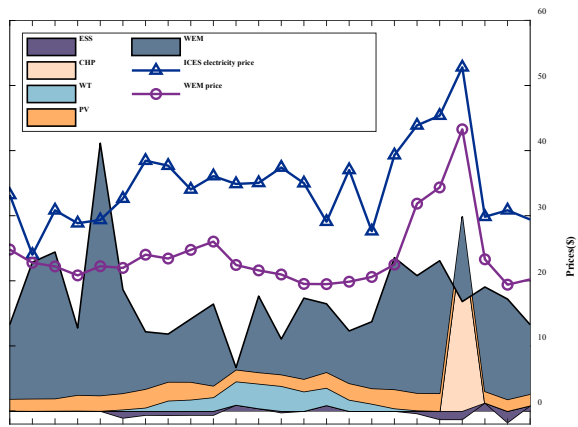

**Fig. 8.** Energy sources and prices for electric power with PD-TD3 method in the summer day

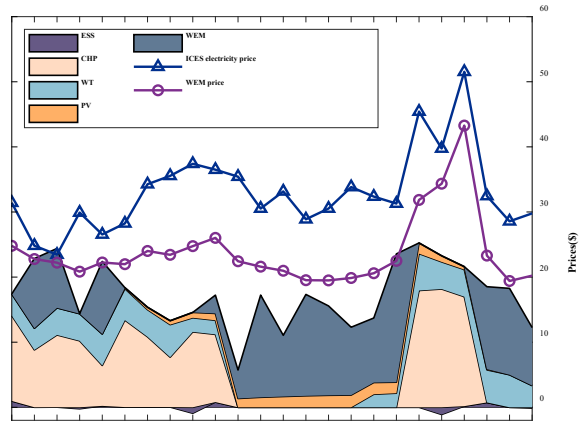

**Fig. 9.** Energy sources and prices for electric power with PD-TD3 method in the winter day

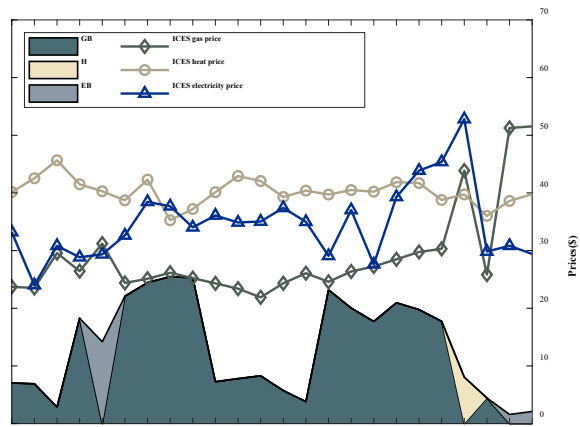

**Fig. 10.** Energy sources and prices for heat power with PD-TD3 method in the summer day

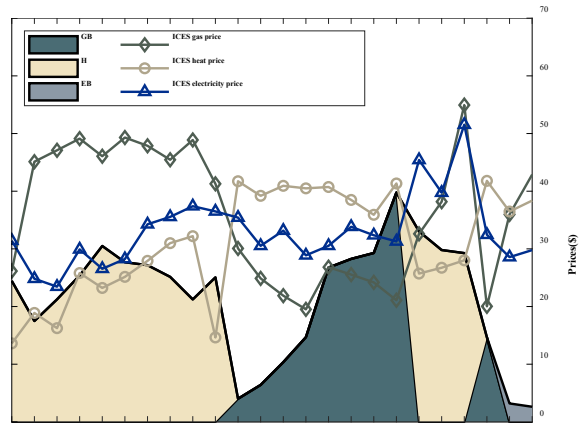

**Fig. 11.** Energy sources and prices for heat power with PD-TD3 method in the winter day

### 5.4 Analysis of pricing and operation decisions

For a more in-depth analysis of the learned pricing and device scheduling policies of the three algorithms above, the energy resources and prices for satisfying power demand and heat demand in the typical test day (the winter scenario) are presented in Figs. 12-13 for comparison of PD-TD3 and S-DDPG. As illustrated in Fig. 8 and Fig. 12, two transaction results show a similar trend in electricity prices in the whole transaction period because of the inherent impact of the same wholesale electricity prices, but the prices in Fig. 8 are higher than those in Fig. 12, therefore resulting in a smoother power consumption curve. The higher prices in Fig. 8 also leads to a low electric power purchase in the wholesale electricity market (WEM) in most periods except 11:00-18:00 due to the low power demand and low prices in WEM. However, the agent of S-DDPG poses a much lower ICES power price, which makes the power consumption curve much steeper. The ICESO of both algorithms determines lower heat prices in periods with a turning on CHP due to its low marginal cost for heat production, while having a lower natural gas price compared to heat in the rest periods.

The devices scheduling is also illustrated in Fig. 12. It can be observed that the PD-TD3 operates CHP for a longer period and purchases less electricity from the WEM compared to S-DDPG. Specifically, the CHP is turned on during the periods of 0:00-11:00 and 18:00-22:00 to sell power and heat to MEUs, since WEM prices, power, and heat demands are relatively higher. In the rest of the hours, the ICESO tends to sell electricity and natural gas from the external market due to the low demand and high cost of the CHP operation. During some periods, the ICESO can provide all the power demand through the power generation from the CHP, DER, and EBS, and the power demand is also cut down or shifted, which is thought of as operating in high energy efficiency. However, the S-DDPG relies on the external market much more by purchasing power and gas to satisfy demand in most periods except for 18:00-22:00, when the CHP is turned on. This is because the S-DDPG algorithm cannot learn the tiny difference in WEM prices and demand between the periods of 0:00-11:00 and 11:00-18:00, even though the market environment during the former periods offers a positive profits uplift for operating the CHP. This not only demonstrates the superior policy of the PD-TD3 compared to S-DDPG but also indicates the high energy efficiency of the MNC-ICES operated by the PD-TD3 algorithm.

Nevertheless, the policies generated by the PD-TD3 and S-DDPG also show differences in physical constraint violations. As the power consumption under PD-TD3 is much smoother and lower compared to the S-DDPG shown in Fig. 12, it is intuitive that the latter may have more constraint violations in the electrical distribution network due to higher power consumption in peak hours (8:00-9:00 and 18:00-21:00). On the other hand, there is a lower value in a single kind of power consumption for satisfying heat demand in a single time slot, which is operated by S-DDPG and shown in Fig. 13. The network safety of gas and heat is easier to guarantee in the former algorithm, while the policy generated by the latter algorithm may transfer the burden of power transmission from the electrical distribution network to the gas and the thermal networks during peak energy consumption periods, which verify the inherent logic of maintaining the operational safety accounting for multi-energy networks.

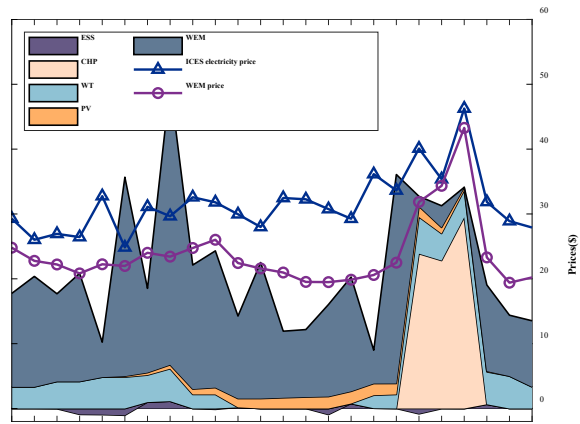

**Fig. 12.** Energy sources and prices for electric power with S-DDPG

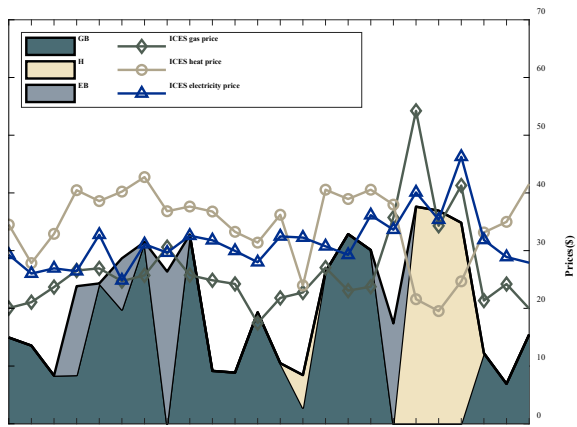

**Fig. 13.** Energy sources and prices for heat power with S-DDPG

## 5.5 Impact of CHP models

As discussed in Section 2 and Appendix, the CHP with a non-convex operating region is effective in generating both electricity and heat, providing system flexibility in reducing the network constraint violation. This subsection examines the influence of the non-convex CHP model on the operation region. For this purpose, a comparison between the scenario with the simplified and detailed CHP model is made, along with the subsequent analysis of the impact on energy transactions and ICES operation. The energy resources and prices in the typical winter day by using simplified CHP are presented in Figs. 14-15. The total reward and network constraint violations with different CHP models are presented in Table 5.

In Fig. 14, CHP's operation periods are cut down from 0:00-7:00 since CHP's power and heat output are constrained, leading to its non-profitability during that period. However, this change also results in higher electricity prices and a different power consumption portfolio with a lower average value in MNC-ICES. It should be noted that the less operational profitability of CHP increases the prices for heat but decreases the gas prices, especially during the 0:00-7:00 in Fig. 15, when non-convex CHP is in operation but simplified one is not, since the ICESO aims to stimulate the MEUs to consume natural gas instead of heat with turned down CHP. Moreover, the heat consumption is affected even in peak hours of heat demand, which is 18:00-22:00. As the heat generation of CHP is constrained to be linearly related to the power generation, and the heat demand is higher than the power demand during 18:00-22:00, the heat generation is limited to a low level, leading to a significant cut-down in heat consumption in Fig. 15 comparing those in Fig. 9. Nevertheless, the implementation of simplified CHP model decreases the total generation of both power and heat, which is shown in Table 5. The generations of power and heat decrease around two times and seven times, respective, while the generation cost only decrease no more than three times since the detailed CHP model has a small marginal generation cost for heat. Finally, the implementation of the simplified model results in a lower cumulative reward of 7698.97, compared to 11593.98 in the detailed CHP model.

Even though the simplified model decreases the profits of ICESO significantly, it results in a smaller physical constraint violation of electricity and heat networks and has a similar violation in terms of gas networks in comparison to the detailed one, and is shown in Table 5. Especially, the cost for heat network violation decreases from 4.91 to 0, because of the significant decrease in heat generation of CHP and the consequent reduced heat consumption of MEUs. On the other hand, due to the substitute effect, MEUs tend to consume more natural gas, putting a burden on the gas network operation. This leads to a slightly higher cost for gas network constraint violations. In summary, the implementation of a simplified model cut down the power and heat generation, and cumulative reward significantly by narrowing the operation region, indicating that the simplified model deviates from the detailed model to a great extent and reflects an unfaithful simulation of the ICES operation in reality.

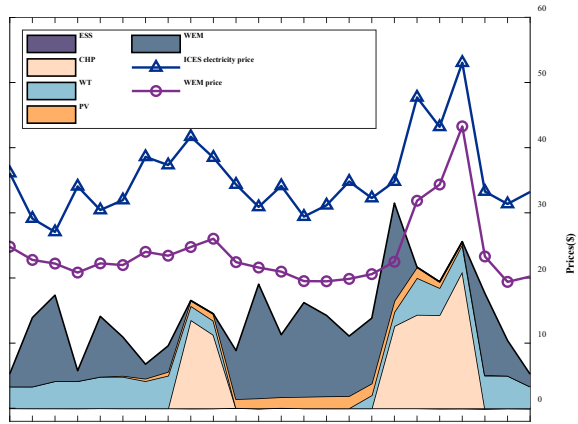

**Fig. 14.** Energy sources and prices for electric power by using simplified CHP

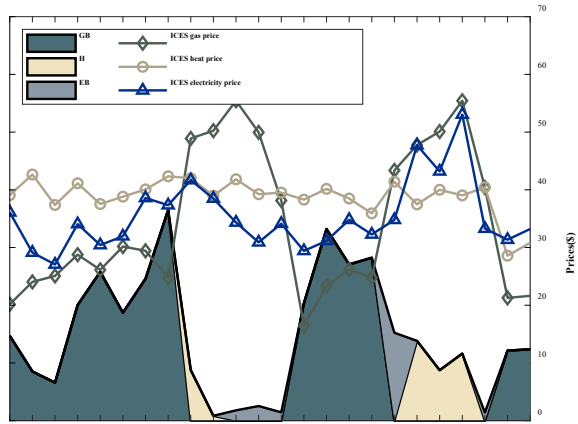

**Fig. 15.** Energy sources and prices for heat power by using simplified CHP

**Table 5**
Results comparison of implementing detailed and simplified CHP models

| | Total reward | Network violation | | | CHP output | | |
|---|---|---|---|---|---|---|---|
| | | E | G | H | E | H | Cost |
| Realistic CHP model with non-convex feasible region | 11593.98 | 0.65 | 0.10 | 4.91 | 163.61 | 501.25 | 5728.65 |
| Simplified CHP model with fixed conversion rate | 7698.97 | 0.56 | 0.20 | 0.00 | 87.04 | 69.63 | 2558.4 |

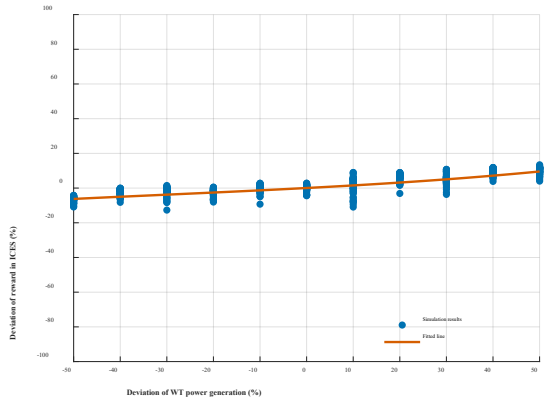

a)

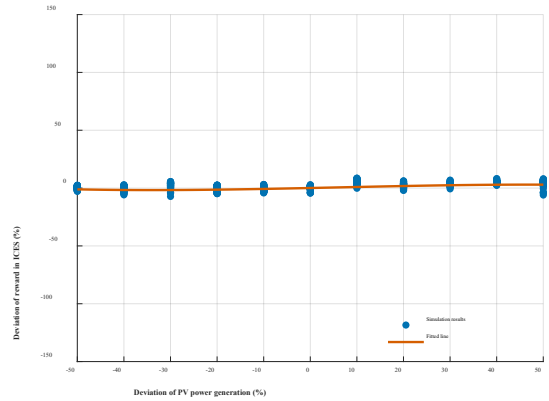

b)

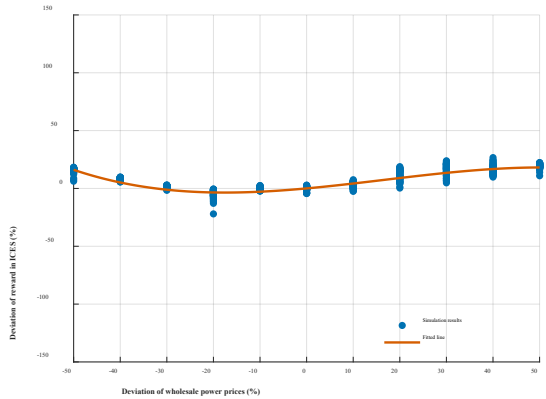

c)

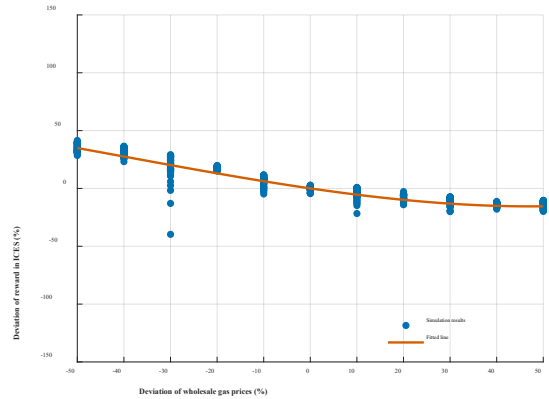

d)

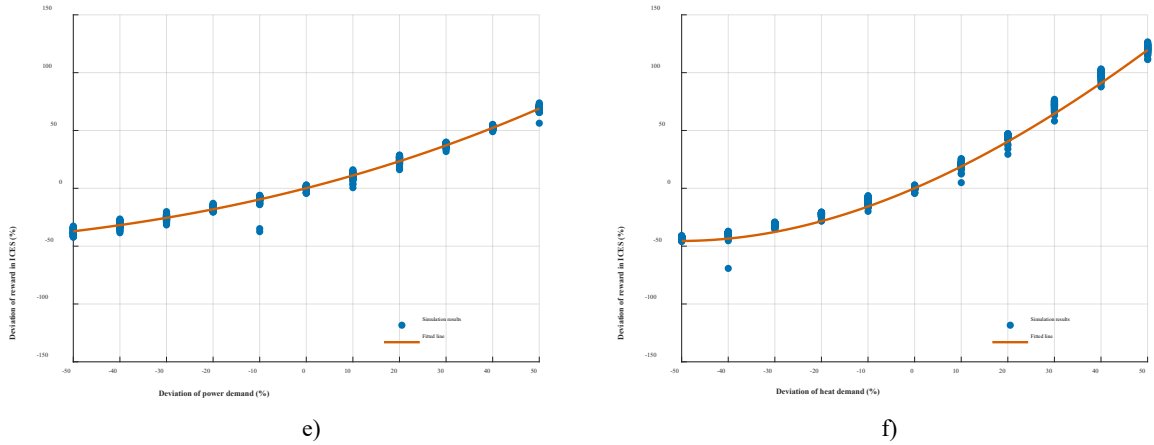

e)

f)

**Fig. 16.** Sensitivity analysis of ICES operation reward on different factors

## 5.6 Sensitivity analysis

In this subsection, we conduct a sensitivity analysis of operational profits (reward) and network constraint violations (cost) to evaluate the impact of various factors on the operational performance of the MNC-ICES model and the proposed Safe RL approach. The tested factors include renewable power generation (wind turbines and photovoltaic systems), wholesale energy prices (electricity and gas), and integrated energy demand levels (electricity and heat), which are considered to introduce the most uncertainties into the MNC-ICES model. Additionally, as the algorithm's performance is significantly influenced by the random seed, which determines the sequence of random numbers generated, the system is simulated 50 times for each scenario using different random seeds. The sensitivities of reward and cost to these factors are evaluated and illustrated in Fig. 16-17, respectively. The horizontal axis represents the variable fluctuation ratio of the factors, ranging from 50% to 150% of the initial configured value in increments of 10%. The vertical axis represents the rate of change in the episodic reward/cost of the ICES. Each data point corresponds to a simulation result for a specific scenario under one random seed with a specific factor adjustment, and the trend line is plotted by fitting to the given data points.

Fig. 16 depicts the sensitivity of reward to changes in various factors. Renewable generation and energy demand positively correlate with reward, exhibiting an approximately linear relationship. Specifically, deviations in energy demand most significantly affect the reward, whereas the reward is least sensitive to renewable generation due to its small proportion in the ICES energy mix. Wholesale gas prices exhibit an approximately linear negative correlation with reward, as higher gas prices increase operational costs. Notably, the wholesale power price shows a nonlinear, likely hyperbolic, relationship with reward. The initial decline in reward corresponds to the natural negative correlation between reward and external energy prices. Conversely, the latter part of the hyperbolic curve is likely due to the increased energy storage arbitrage opportunities created by larger wholesale price gaps.

Fig. 17 illustrates the sensitivity of cost to changes in various factors. Among these, renewable generation has the weakest negative correlation with cost, with its impact being almost negligible. Energy prices demonstrate a certain hyperbolic relationship: initially, an increase in the price of a single energy source decreases costs for a network with lower energy consumption, while an increase in energy price improves the consumption of alternative energies, thereby raising network constraint violations for other energies. Fluctuations of energy demands show a piecewise linear relationship to the cost; energy demand below a certain level results in almost zero network violations, whereas demand above this threshold leads to linear growth in network constraints. Notably, thermal energy demand and wholesale power prices impact network constraint costs most. In contrast, wholesale gas prices and power demand have a weaker impact. The influence of PV and WT is relatively minor and can almost be disregarded.

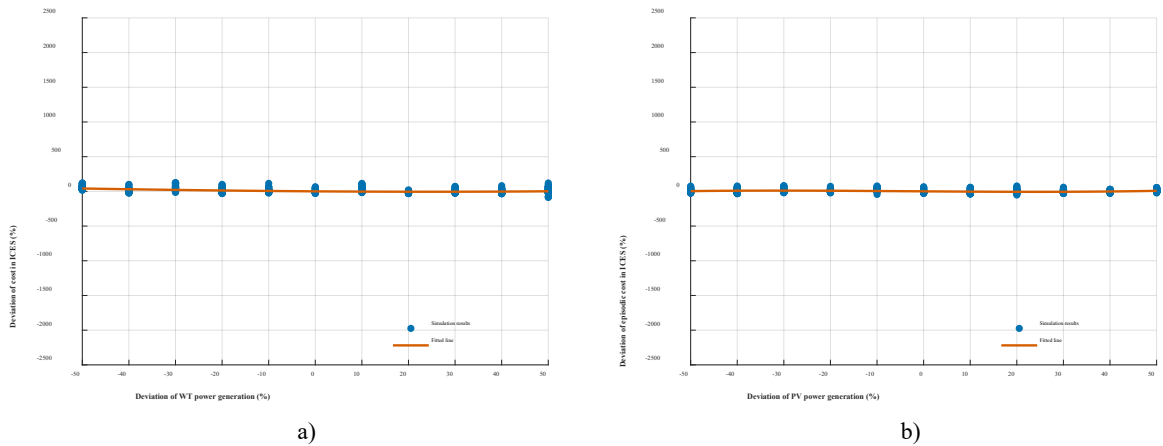

a)

b)

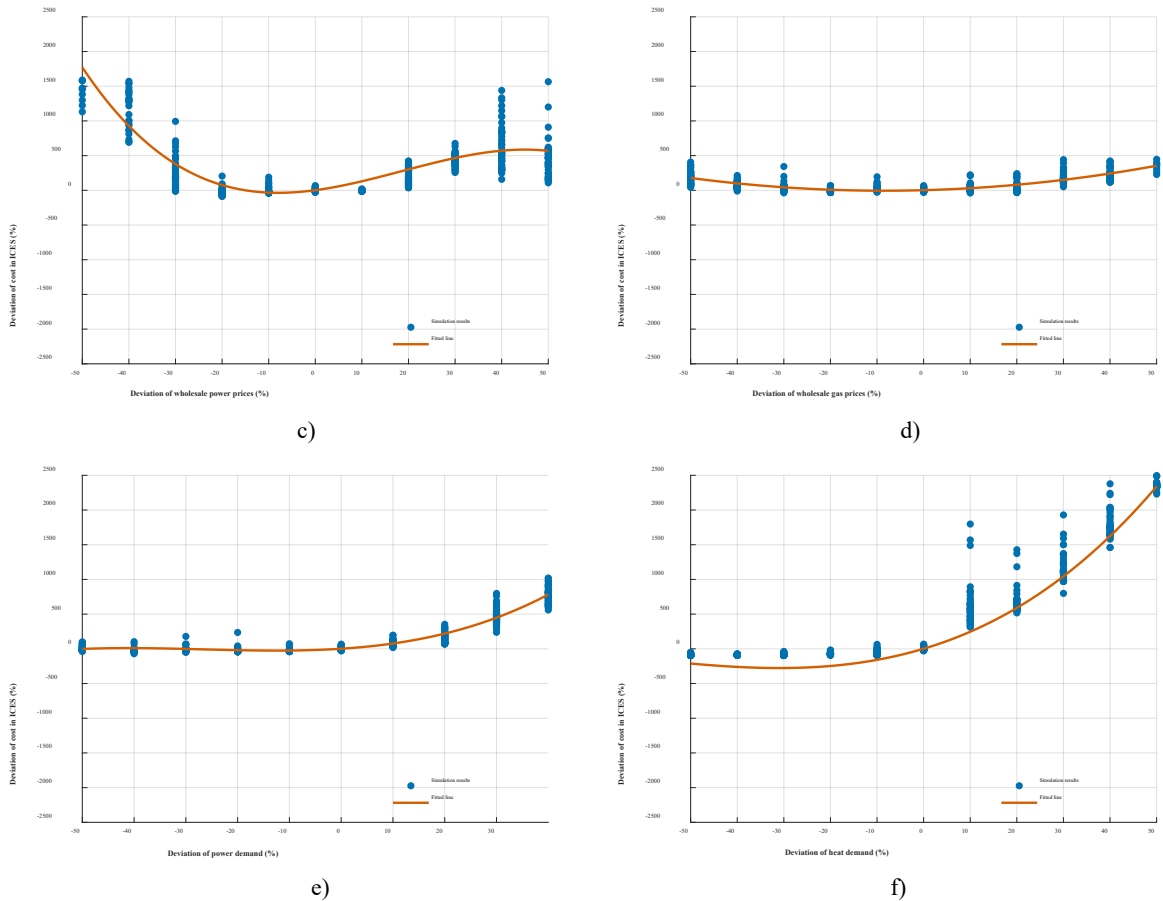

c)                                                                    d)

e)                                                                    f)

**Fig. 17.** Sensitivity analysis of ICES operation network constraints violation (cost) on different factors

### 5.7 Impact of hyperparameters

As hyperparameters have a great impact on algorithm performance, sensitivity analysis is conducted on selected hyperparameters, mainly including the Q-network learning rate and actor-network learning rate. Fig. 18 shows the evolution of cumulative reward and cost under different actor-network (policy) learning rates. It can be observed in Fig. 18 a) that the episode reward can converge to a high value fast with a learning rate lying from 1e-4 to 5e-4 but may converge to a low value, which is a local optimal, in several episodes with a policy learning rate from 6e-4. Also, the curve of the cumulative reward increases faster in the initial stage with a lower learning rate in policy, which means the exploration in the initial stage contains a higher proportion of useless stochastic noise compared to the later stage. Among these parameter settings, the policy learning rate of 4e-4 (purple) can assist in achieving the highest accumulative reward. When comparing Fig. 18 a)-b), the policy learning rate with a higher cumulative reward always results in a higher cost for constraint violation. This demonstrates that the converged cumulative reward has a positive correlation with the cumulative cost, while the policy learning rate plays a key role in the tradeoff of the reward and cost. As all of these parameter settings satisfy the safe operating range of 0~10, the policy learning rate with the highest cumulative reward, which is 4e-4, is selected to be the final setting of the algorithm.

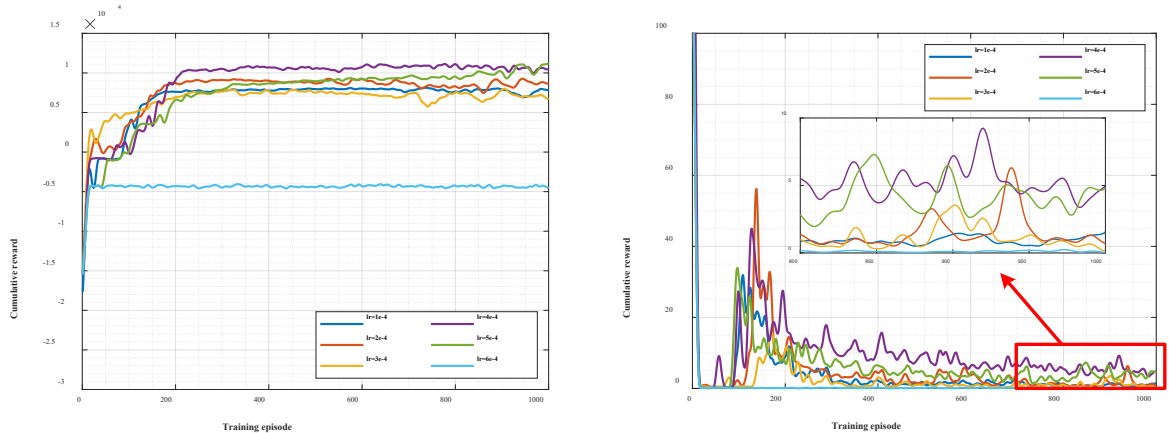

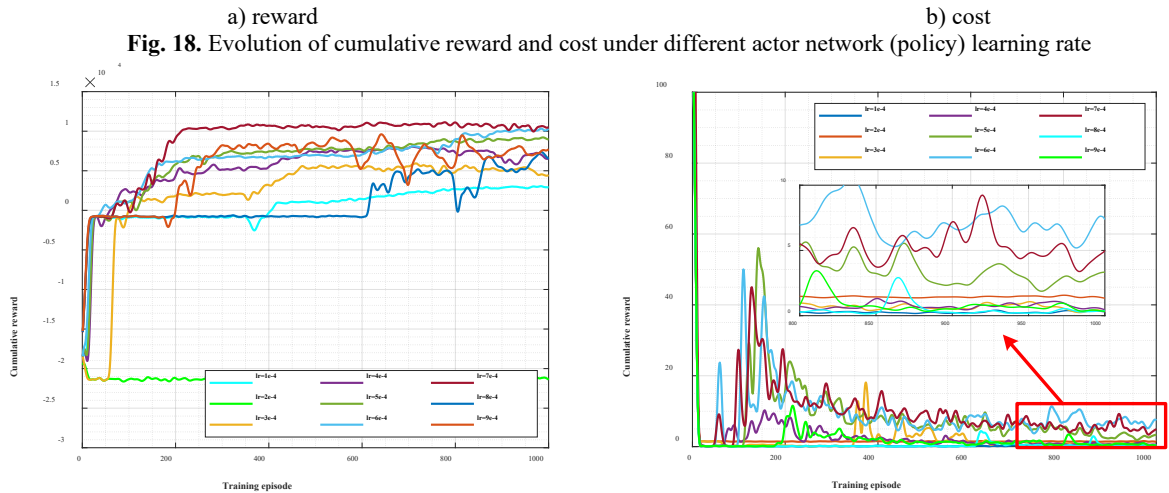

a) reward　　　　　　　　　　　　　　　　　　b) cost

**Fig. 18.** Evolution of cumulative reward and cost under different actor network (policy) learning rate

a) reward　　　　　　　　　　　　　　　　　　b) cost

**Fig. 19.** Evolution of cumulative reward and cost under different critic network (Q-value) learning rate

As for the critic network learning rate shown in Fig. 19, the converged reward, as well as the cost, firstly increases and then decreases with the growing actor-network learning rate, and the learning rate of 7e-4 shows the highest cumulative reward. In general, the evolution curve with a lower learning rate tends to increase gently, while it shows a steep increase or decrease in reward and cost with a higher learning rate. Moreover, a positive correlation between the reward and cost can also be observed for different settings in Fig. 19. Interestingly, the cumulative cost with the learning rate of 5e-4 and 6e-4 may exceed the tolerated cost during the training process, while the cost curve with 8e-4 is nearly zero and is considered too conservative. Among these parameter settings of the critic learning rate, the settings of 7e-4 can balance the reward and cost and increase the algorithm performance to the greatest extent.

## 6. Conclusion

In conclusion, this paper proposes an MNC-ICES model to describe community-level energy systems. The proposed model comprehensively modelled multi-network for integrated energy, realistic energy devices, renewable uncertainty, and IDR of MEUs. Within the community, the ICESO schedules energy devices and prices integrated energy to maximize operational profits while securing the system operation within the safety requirements imposed by integrated network constraints. It provides a basis for practical network-constrained community operation tools and can be referred for software development in energy system operation. Numerical results reveal that the realistic model significantly differs and can attain a higher economic value than simplified models. A novel Safe RL algorithm, PD-TD3, is developed to solve the constrained optimization problem in MNC-ICES and learn the optimal scheduling strategies to maximize profits without violating safety constraints dramatically. The proposed algorithm is based on the Lagrangian method, utilizing a Lagrangian multiplier to penalize the constraint violation during the policy updates. Double networks are employed to mitigate the Q value over-estimation issue of both reward and cost, enabling accurate updates of the Lagrangian multiplier and achieving a balanced tradeoff between the reward and cost. The simulation results demonstrate the superior computational performance and the optimality of the proposed algorithm compared with several benchmarks. Finally, the sensitivity of the MNC-ICES models and the proposed algorithm on model factors and hyperparameters are also analyzed. This work is impactful with potential beneficiaries, including ICES operators and residents, as well as reinforcement learning researchers and practitioners.

Future work will extend this research to long-term strategic planning regarding low-carbon dispatch. Specifically, the strategic planning problem (placing and sizing distributed energy resources) and low carbon dispatch, considering carbon transactions, will be jointly considered in the proposed model. The influence of planning problems on carbon emission accounting for multi-network constraints is an exciting topic and needs further research. Moreover, various types of loads, including air conditioning, electric vehicles, etc., have a significant impact on the ICES operation due to their distinct operational characteristics. Developing operational model and optimization methods accounting for multiple types of loads are meaningful to further improve the community demand flexibility and also stability. Additionally, the multi-energy transactions among several communities are strategic and complex. The equilibrium may share more insights into energy market mechanisms in distribution systems, which is also a potential topic in future research.

## Appendix

Combined Heat and Power Unit

The FOR of the CHP is divided into two convex sections and is represented as follows [19].

$$P_{CHP.n}^t - P_{CHP.n}^B - \frac{P_{CHP.n}^B - P_{CHP.n}^C}{H_{CHP.n}^B - H_{CHP.n}^C} \times \left(H_{CHP.n}^t - H_{CHP.n}^B\right) \leq 0, \forall t \in T \tag{A.1}$$

$$P_{CHP}^t - P_{CHP}^C - \frac{P_{CHP}^C - P_{CHP}^D}{H_{CHP}^C - H_{CHP}^D} \times \left(H_{CHP}^t - H_{CHP}^C\right) \leq 0, \forall t \in T \tag{A.2}$$

$$-\left(1 - \overline{X}_{CHP}^t\right) \times \Gamma \leq P_{CHP}^t - P_{CHP}^E - \frac{P_{CHP}^E - P_{CHP}^F}{H_{CHP}^E - H_{CHP}^F} \times \left(H_{CHP}^t - H_{CHP}^E\right), \forall t \in T \tag{A.3}$$

$$-\left(1 - \underline{X}_{CHP}^t\right) \times \Gamma \leq P_{CHP}^t - P_{CHP}^D - \frac{P_{CHP}^D - P_{CHP}^E}{H_{CHP}^D - H_{CHP}^E} \times \left(H_{CHP}^t - H_{CHP}^D\right), \forall t \in T \tag{A.4}$$

$$\overline{X}_{CHP}^t + \underline{X}_{CHP}^t = I_{CHP}^t, \forall t \in T \tag{A.5}$$

$$-\left(1 - \underline{X}_{CHP}^t\right) \times \Gamma \leq H_{CHP}^t - H_{CHP}^E \leq \left(1 - \overline{X}_{CHP}^t\right) \times \Gamma, \forall t \in T \tag{A.6}$$

$$0 \leq P_{CHP}^t \leq P_{CHP}^A \times I_{CHP}^t, \forall t \in T \tag{A.7}$$

$$0 \leq H_{CHP}^t \leq H_{CHP}^A \times I_{CHP}^t, \forall t \in T \tag{A.8}$$

In equations above, $P_{CHP}^t, H_{CHP}^t$ are generated power and heat for the CHP in time slot t. As the region is described by a non-convex polygon, $P_{CHP}^A$ and $H_{CHP}^A$ indicate the power and heat output of the CHP at point $A$ in the feasible region, and the same applied to the other points $BCDEF$. $\overline{X}_{CHP}^t(\underline{X}_{CHP}^t)$ states the operating status in the first (second) convex section, when the CHP operate in the first (second) section, $\overline{X}_{CHP}^t(\underline{X}_{CHP}^t) = 1$, and $\underline{X}_{CHP}^t\left(\overline{X}_{CHP}^t\right) = 0$. $\Gamma$ denotes a sufficiently large number to assist model description, while $I_{CHP}^t$ is the commitment status of the CHP. The total operation cost of the CHP unit at time t can be expressed by equation (A.9), where $a_{CHP}$, $b_{CHP}$, $c_{CHP}$, $d_{CHP}$, $e_{CHP}$ and $f_{CHP}$ represent the cost coefficients.

$$C_{CHP}^t(P_{CHP}^t, H_{CHP}^t) = a_{CHP}P_{CHP}^{t\ 2} + b_{CHP}P_{CHP}^t + c_{CHP} + d_{CHP}H_{CHP}^{t\ 2} + e_{CHP}H_{CHP}^t + f_{CHP}P_{CHP}^t H_{CHP}^t \tag{A.9}$$

### Distributed Energy Resources

As VRE, wind power inherently carries high uncertainty. The wind speed ($\omega$), directly influencing power output, is predicted with an unavoidable error $\Delta\omega$ in (A.10), which is modelled by a Weibull PDF [48]. The power output $P_{WT}^t$ of WT is modeled in equation (A.11), where it is positive if and only if the wind speed exceeds the starting speed ($\omega_{in}^c$); otherwise, $P_{WT}^t$ is always zero. The upper limit for WT power is $P_{WT.rated}^t$ when $\omega_{rated}^c \leq \omega \leq \omega_{out}^c$. If the wind speed surpasses the cutout speed $\omega_{out}^c$, WT will be cut out, resulting in $P_{WT} = 0$. Additionally, the Weibull PDF is employed to estimate the uncertainty parameter due to wind speed prediction errors in (A.12)

$$\omega = \omega_{fs} + \Delta\omega \tag{A.10}$$

$$P_{WT}^t(\omega) = \begin{cases} 0, & \omega \leq \omega_{in}^c \quad or \quad \omega \geq \omega_{out}^c \\ \dfrac{\omega + \omega_{in}^c}{\omega_{rated} + \omega_{in}^c} P_{WT.rated}^t, & \omega_{in}^c \leq \omega \leq \omega_{rated}^c \\ P_{WT.rated}^t, & \omega_{rated}^c \leq \omega \leq \omega_{out}^c \end{cases} \tag{A.11}$$

$$F_\omega(\Delta\omega; \lambda, k) = \begin{cases} \dfrac{k}{\lambda}\left(\dfrac{\Delta\omega + 0.5}{\lambda}\right)^{k-1} e^{-\left(\frac{\Delta\omega + 0.5}{\lambda}\right)^k} & \Delta\omega \geq -0.5 \\ 0, & \Delta\omega < -0.5 \end{cases} \tag{A.12}$$

For photovoltaic power generation, the prediction error $\Delta I$ of PV is introduced in (A.13). PV generates electricity by converting solar radiation energy, and power generation is directly related to solar irradiance in (A.14). The Beta PDF is employed to estimate uncertain parameters with minimal error in (A.15).

$$I = I_{fs} + \Delta I \tag{A.13}$$

$$P_{PV}^t = \sum_{n \in N_{pv}} \eta_{pv_n} S_{pv_n} I^t \tag{A.14}$$

$$F_S(\Delta I; \alpha, \beta) = \frac{(\Delta I + 0.5)^{\alpha-1}\left(1 - (\Delta I + 0.5)\right)^{\beta-1}}{\int_0^1 u^{\alpha-1}(1-u)^{\beta-1}du} \tag{A.15}$$

### Energy Storage Systems

The detailed model of EBS is shown as follows.

$$E_{EBS}^t = (1 - \beta_{EBS})E_{EBS}^{t-1} + P_{EBS.c}^t \eta_{EBS.c} - P_{EBS.d}^t \tag{A.16}$$

$$0 \leq P_{EBS.c}^t \leq S_{EBS.c}^t P_{EBS.max} \tag{A.17}$$

$$0 \leq P_{EBS.d}^t \leq S_{EBS.d}^t P_{EBS.max} \tag{A.18}$$

$$S_{EBS.c}^t + S_{EBS.d}^t \leq 1 \tag{A.19}$$

$$0 \leq E_{EBS}^t \leq E_{EBS.max} \tag{A.20}$$

In equations above, $E_{EBS}^t$ is the battery capacity at time interval $t$. $\beta$ and $\eta_{EBS.c}$ are predetermined parameters representing the loss factor and charging efficiency, respectively. $P_{EBS.c}^t$ and $P_{EBS.d}^t$ represents the charging power and discharging power at time step $t$, respectively. $S_{EBS.c}^t$ and $S_{EBS.d}^t$ represent the charging state and discharging state at time step $t$, respectively. $P_{EBS.c.max}$ and $P_{EBS.d.max}$ are the maximum charging and discharging power, respectively. $E_{EBS.min}$ and $E_{EBS.max}$ represent the upper and lower limits of battery capacity, respectively.

In the model above, the representation of the state of charge (SoC) is shown in (A.16). The maximal charge and discharge power are constrained by (A.17) and (A.18), respectively. (A.19) is employed to determine the charge of discharge state of EBS. (A.20) constrains the range of total capacity of the energy in EBS.

A generalized energy storage system model is applied to address TES. This model aligns with that of EBS and is not detailed here for the sake of brevity.

$$E_{TES}^t = (1 - \beta)E_{TES}^{t-1} + H_{TES.c}^t \eta_{TES.c} - H_{TES.d}^t \tag{A.21}$$

$$0 \leq H_{TES.c}^t \leq S_{TES.c}^t H_{TES.max} \tag{A.22}$$

$$0 \leq H_{TES.d}^t \leq S_{TES.d}^t H_{TES.max} \tag{A.23}$$

$$S_{TES.c}^t + S_{TES.d}^t \leq 1 \tag{A.24}$$

$$0 \leq E_{TES}^t \leq E_{TES.max} \tag{A.25}$$

Multiple Energies Users

$$E_{MEU}\left(P_{i,e}^t\right) = \begin{cases} \omega_i^t - \dfrac{\lambda_i^t}{2}\left(P_{i,e}^t\right)^2, 0 \leq P_{i,e}^t \leq \dfrac{\omega_i^t}{\lambda_i^t} \\ \dfrac{(\omega_n^t)^2}{2\lambda_i}, P_{i,e}^t > \dfrac{\omega_i^t}{\lambda_i} \end{cases} \tag{A.26}$$

$$H_{MEU}\left(H_{i,eb}^t, H_{i,gb}^t, H_{i,h}^t\right) = -\sigma_i^t\left(H_{i,eb}^t + H_{i,gb}^t + H_{i,h}^t\right)^2 + \varsigma_i^t\left(H_{i,eb}^t + H_{i,gb}^t + H_{i,h}^t\right) \tag{A.27}$$

In (A.26), $P_{i,e}^t$ is the power consumption of MEU $i$ during time interval $t$. $\omega_i$ and $\lambda_i$ are preset parameters reflecting the preference of MEU in energy consumption. In (A.27), $H_{i,eb}^t$, $H_{i,gb}^t$ and $H_{i,h}^t$ are heat power from EB, GB, and ICESO, respectively. Similarly, $\sigma_i^t$ and $\varsigma_i^t$ are preset parameters. By considering the utility above, the objective function of MEUs is modelled by (A.28) and is constrained by (A.29)-(A.33).

$$\max_{P_{i,e}^t, P_{i,eb}^t, P_{i,gb}^t, H_{i,h}^t} U_{MEU} = \sum_{t=1}^{T} \left\{ \underbrace{\left(E_{MEU}\left(P_{i,e}^t\right) + H_{MEU}\left(H_{i,eb}^t, H_{i,gb}^t, H_{i,h}^t\right)\right)}_{\text{Utility for energy consumption}} - \underbrace{\left(x_{r,e}^t\left(P_{i,e}^t + P_{i,eb}^t\right) + x_{r,g}^t P_{i,gb}^t + x_{r,h}^t P_{i,h}^t\right)}_{\text{Cost for energy purchase}} \right\} \tag{A.28}$$

$s.t.$

$$H_{i,eb}^t = \eta_{EB,i} P_{i,eb}^t \tag{A.29}$$

$$H_{i,gb}^t = \eta_{GB,i} P_{i,gb}^t \tag{A.30}$$

$$0 \leq P_{i,e}^t \leq P_{i,e.max}^t \tag{A.31}$$

$$0 \leq P_{i,eb}^t \leq P_{i,eb.max}^t \tag{A.32}$$

$$0 \leq P_{i,gb}^t \leq P_{i,gb.max}^t \tag{A.33}$$

In these equations, $P_{i,eb}^t$ and $P_{i,gb}^t$ are power and gas consumed by EB and GB of MEU $i$ during time interval $t$, respectively. $\eta_{EB,i}$ and $\eta_{GB,i}$ are the energy conversion rates of EB and GB for MEU $i$. $P_{i,e.max}^t$, $P_{i,eb.max}^t$ and $P_{i,gb.max}^t$ are the upper bounds for the corresponding power consumption of electric appliance, power consumption of EB, and gas consumption of GB. (A.28) is the objective function of the MEU with the decision variables of $P_{i,e}^t, P_{i,eb}^t, P_{i,gb}^t, H_{i,h}^t$. In (A.28), the first term is the utility for integrated energy consumption, and the second term is the cost for integrated energy purchase from ICESO. Equations (A.29) and (A.30) are the equality constraints for the power conversion of EB and GB. Inequalities (A.31)-(A.33) are constraints for electricity consumption, and power input for EB and GB, respectively.